\documentclass[sigconf]{acmart}

\AtBeginDocument{%
  \providecommand\BibTeX{{%
    \normalfont B\kern-0.5em{\scshape i\kern-0.25em b}\kern-0.8em\TeX}}}

\usepackage{bbding}
\usepackage{subfigure}
\usepackage[boxed,ruled,lined]{algorithm2e}
\usepackage{multirow}
\usepackage{makecell}

\usepackage{arydshln}

\newcommand{\paratitle}[1]{\vspace{0.8ex}\noindent\textbf{#1}}
\newcommand{\ourname}{UniSAR\xspace}

\newcommand{\rtr}{\textbf{r2r}\xspace}
\newcommand{\str}{\textbf{s2r}\xspace}
\newcommand{\rts}{\textbf{r2s}\xspace}
\newcommand{\sts}{\textbf{s2s}\xspace}

\copyrightyear{2024} 
\acmYear{2024} 
\setcopyright{acmlicensed}
\acmConference[SIGIR '24]{Proceedings of the 47th International ACM SIGIR Conference on Research and Development in Information Retrieval}{July 14--18, 2024}{Washington, DC, USA}
\acmBooktitle{Proceedings of the 47th International ACM SIGIR Conference on Research and Development in Information Retrieval (SIGIR '24), July 14--18, 2024, Washington, DC, USA}
\acmDOI{10.1145/3626772.3657811}
\acmISBN{979-8-4007-0431-4/24/07}

\begin{document}

\title{\ourname: Modeling User Transition Behaviors between\\ Search and Recommendation}

\author{Teng Shi}
\affiliation{%
\institution{Renmin University of China}
  \city{Beijing}\country{China}
  }
\email{shiteng@ruc.edu.cn}

\author{Zihua Si}
\affiliation{
\institution{Renmin University of China}
  \city{Beijing}\country{China}
  }
\email{zihua_si@ruc.edu.cn}

\author{Jun Xu}
\authornote{Corresponding authors. Work partially done at Engineering Research Center of Next-Generation Intelligent Search and Recommendation, Ministry of Education. \\
Work done when Teng Shi and Zihua Si were interns at Kuaishou.
}
\affiliation{
  \institution{Renmin University of China}
  \city{Beijing}\country{China}
  }
\email{junxu@ruc.edu.cn}

\author{Xiao Zhang}
\affiliation{
  \institution{Renmin University of China}
  \city{Beijing}\country{China}
  }
\email{zhangx89@ruc.edu.cn}

\author{Xiaoxue Zang}
\affiliation{
  \institution{Kuaishou Technology Co., Ltd.}
  \city{Beijing}\country{China}
  }
\email{zangxiaoxue@kuaishou.com}

\author{Kai Zheng}
\affiliation{
  \institution{Kuaishou Technology Co., Ltd.}
  \city{Beijing}\country{China}
  }
\email{zhengkai@kuaishou.com}

\author{Dewei Leng}
\affiliation{
  \institution{Kuaishou Technology Co., Ltd.}
  \city{Beijing}\country{China}
  }
\email{lengdewei@kuaishou.com}

\author{Yanan Niu}
\affiliation{
  \institution{Kuaishou Technology Co., Ltd.}
  \city{Beijing}\country{China}
  }
\email{niuyanan@kuaishou.com}

\author{Yang Song}
\affiliation{
  \institution{Kuaishou Technology Co., Ltd.}
  \city{Beijing}\country{China}
  }
\email{yangsong@kuaishou.com}

\renewcommand{\shortauthors}{Teng Shi et al.}

\begin{abstract}
Nowadays, many platforms provide users with both search and recommendation services as important tools for accessing information. The phenomenon has led to a correlation between user search and recommendation behaviors, providing an opportunity to model user interests in a fine-grained way. 
Existing approaches either model user search and recommendation behaviors separately or 
overlook the different transitions between user search and recommendation behaviors.
In this paper, we propose a framework named \textbf{\ourname} that effectively models the different types of fine-grained behavior transitions for providing users a \textbf{Uni}fied 
\textbf{S}earch \textbf{A}nd \textbf{R}ecommendation service. 
Specifically, UniSAR models the user transition behaviors between search and recommendation through
three steps: extraction, alignment, and fusion, which are respectively implemented by transformers equipped with pre-defined masks, contrastive learning that aligns the extracted fine-grained user transitions, and cross-attentions that fuse different transitions. To provide users with a unified service, the learned representations are fed into the downstream search and recommendation models. Joint learning on both search and recommendation data is employed to utilize the knowledge and enhance each other. 
Experimental results on two public datasets demonstrated the effectiveness of \ourname in terms of enhancing both search and recommendation simultaneously. 
The experimental analysis further validates that \ourname enhances the results by successfully modeling the user transition behaviors between search and recommendation.
\end{abstract}



\maketitle

\section{Introduction}

Search engines and recommender systems have become important tools for users to access information.
Many commercial platforms, such as YouTube, TikTok, and Kwai, have provided both search and recommendation (\textbf{S\&R}) services in one app, blending the boundaries between users' S\&R behaviors.
In these apps, users may further search for related items after browsing their preferred items, or they might explore similar items following a search for a specific item.
These phenomena indicate that understanding the user behavior transitions between S\&R services is highly beneficial for modeling user interests.

Early studies like JSR~\cite{JSR,JSR2} optimize S\&R models by a joint training loss, where both models share item representations. Recently, researchers have paid attention to modeling user behaviors in two scenarios.
SESRec~\cite{SESRec} and UnifiedSSR~\cite{xie2023unifiedssr} proposed to capture user interests from S\&R behaviors separately with distinct encoders.
Meanwhile, USER~\cite{USER} proposed to mix the S\&R histories into one sequence according to timestamp and leverage a single~encoder.

\begin{figure}[t]
    \centering
    \includegraphics[width=0.85\columnwidth]{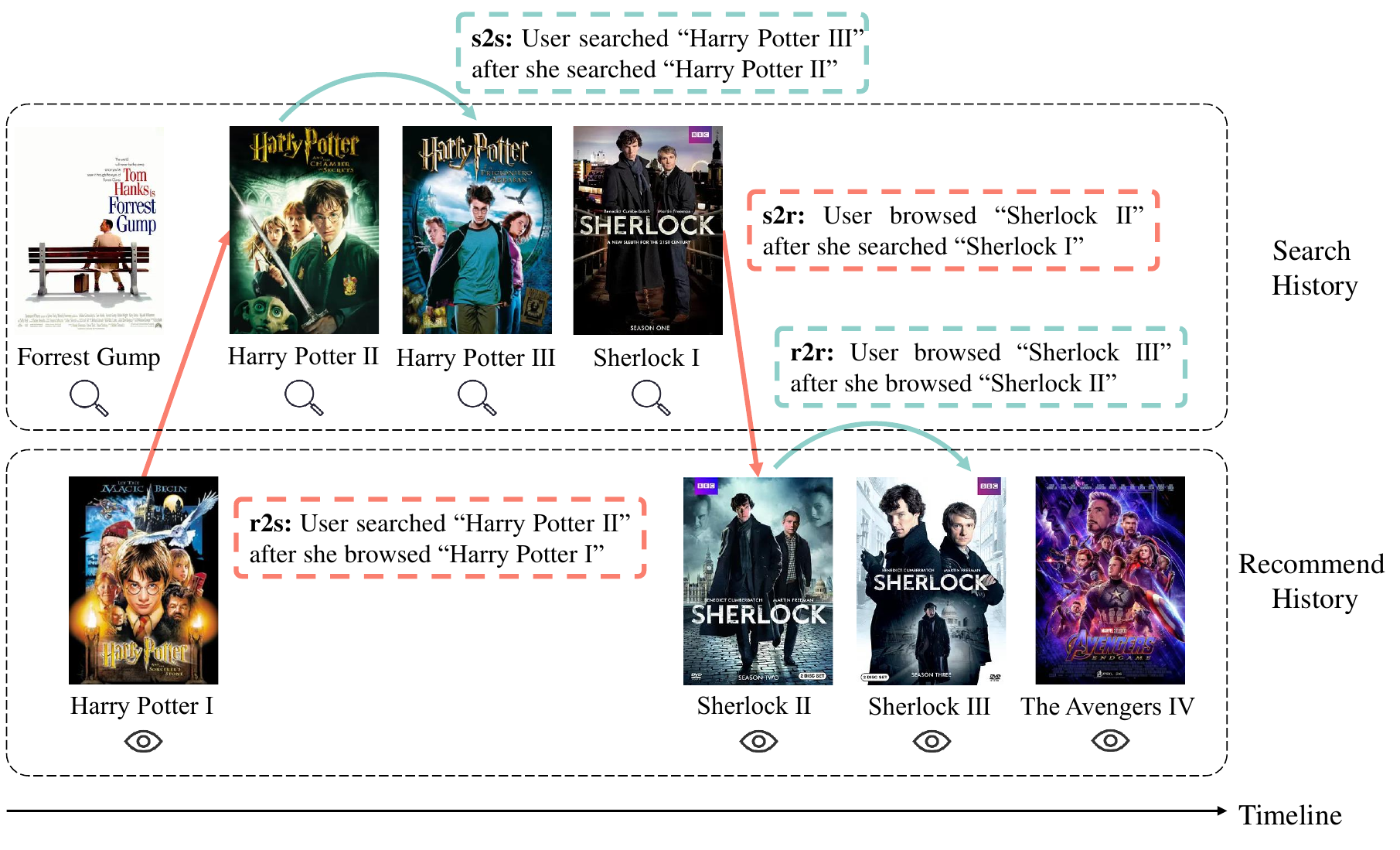}
    \vspace{-8px}
    \caption{
    An example of the user search and recommendation history in an App with both search and recommendation services. The user exhibits four types of transition behaviors, as shown by four arrows.}
    \label{fig:introduction-example}
    \vspace{-0.5cm}
\end{figure}

Despite their effectiveness, existing studies have no mechanism to handle the fine-grained user transitions between S\&R explicitly.
The mixture of search and recommendation scenarios in one app makes user behaviors much more complex than those of a single search engine or recommender system. For example, users may switch to search/recommendation services if the current recommendation/search results cannot satisfy their needs well, leading to a large number of transition behaviors. Specifically, as illustrated in Figure~\ref{fig:introduction-example}, a typical user may exhibit four types of behaviors: (1) staying in the search scenario and issuing new queries (denoted as \textbf{s2s}); (2) keeping browsing new items in the recommendation scenario (denoted as \textbf{r2r}); 
(3) searching for certain items after browsing recommended ones, driven by new information needs 
(transits from recommendation to search, denoted as \textbf{r2s}); 
and (4) browsing the recommended items after searching, as they align well with user interests (transits from search to recommendation, denoted as \textbf{s2r}).

\begin{figure}
    \centering
    \includegraphics[width=0.95\columnwidth]{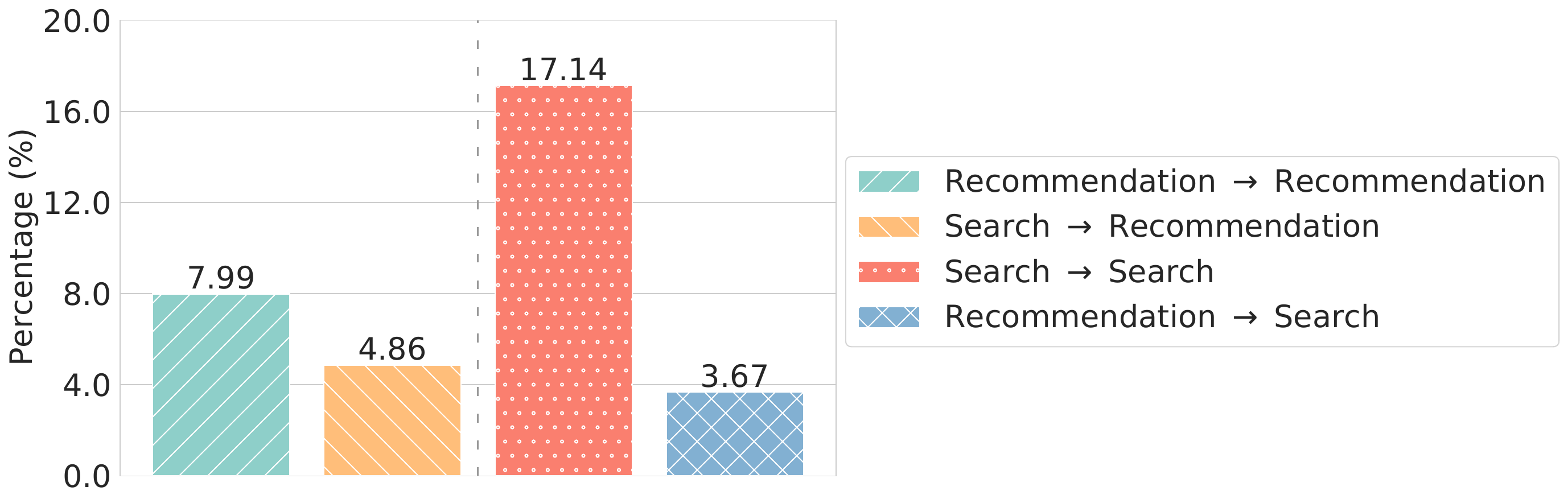}
    \vspace{-0.3cm}
    \caption{The percentage of the immediately preceding clicked item correlates with the current item.
    }
    \label{fig:introduction-analysis}
    \vspace{-0.6cm}
\end{figure}

We conducted an analysis based on KuaiSAR~\cite{Sun2023KuaiSAR} to show whether the information needs of a user also changed when she transits between search and recommendation. For example, given a randomly sampled set of recommended and clicked items (current item), we calculated the percentage that the immediate preceding clicked item is correlated\footnote{
Two items are considered correlated if they are within the same category.}, grouped by the preceding scenario (search or recommendation).
From the results of the left two columns of Figure~\ref{fig:introduction-analysis}, we can see that if the user's preceding scenario is also the recommendation (no scenario transition), the correlated percentage is 7.99\%. However, if the user's preceding scenario is search (transit from search to recommendation), the correlated percentage drops to 4.86\%. Similarly, given a clicked item in search, we also calculated the aforementioned two percentages, shown in the right two columns of Figure~\ref{fig:introduction-analysis}. We can see that if the user preceding scenario is the same (i.e., search), the correlated percentage is 17.14\%. The number drops to 3.67\% if the preceding scenario is different (i.e., recommendation). From the analysis, we conclude that it has much more possibility that a user has some new information needs if she transits between S\&R. The phenomenon motivates the necessity of fine-grained modeling of user transition behaviors.

This paper proposes an approach named \textbf{\ourname} which explicitly models different types of fine-grained user transition behaviors for both S\&R.
\ourname primarily models user transitions through three components: \emph{extraction}, \emph{alignment}, and \emph{fusion}.
(1)~In the extraction, the chronologically sorted S\&R history is fed into a transformer equipped with the mask mechanism.
The mask mechanism enables attention calculations between different behaviors, aiding in the extraction of \rts and \str transitions.
Additionally, the S\&R histories are separately inputted into two other transformers to extract \sts and \rtr transitions.
(2)~The alignment aims to align transitions from the same scenarios with those from different scenarios (more likely indicating new information needs) to enable the model to learn relationships between different transitions. 
Specifically, contrastive learning is employed to align \rts with \sts and \str with \rtr, enabling the model to comprehend relationships between them.
(3)~The fusion utilizes the cross-attention to fuse \rts and \sts, resulting in the representation of search. At the same time, another cross-attention is used to fuse \str and \rtr, achieving the representation of recommendation.
The two representations constitute the overall user representation.

These representations, along with candidate items and queries, can be used to predict user preferences, as have been used in traditional search and recommendation models. As an example, we jointly train \ourname on S\&R data, enabling its application in both scenarios.
MMoE~\cite{ma2018modeling,tang2020progressive} is employed to alleviate the seesaw phenomenon for multi-task training.

The major contributions of the paper are summarized as follows:

\noindent\textbf{$\bullet $}~\ourname proposes three components of extraction, alignment, and fusion, to effectively capture the fine-grained transitions between users' S\&R behaviors.
We leverage different transformers to extract four transitions and contrastive learning techniques to align them. Cross-attention mechanisms are further used to fuse different~transitions.

\noindent\textbf{$\bullet $}~\ourname is trained jointly on search and recommendation tasks and can be applied to both scenarios, effectively utilizing knowledge from each to enhance the other.

\noindent\textbf{$\bullet $}~Experiment results on two public datasets demonstrate the effectiveness of \ourname.
\ourname has outperformed not just the traditional models for a single scenario but also surpassed the existing joint search and recommendation models, achieving state-of-the-art (SOTA) performance.
\section{Related Work}

\paratitle{Sequential Recommendation.}
Sequential recommendation aims to model user historical behaviors to capture users' dynamic preferences. 
Early works~\cite{rendle2010factorizing} adopt the Markov Chain to learn item transition probability.
With the advancement of deep learning, various works have been proposed to model user historical behaviors using neural networks, including methods based on Recurrent Neural Network (RNN)~\cite{GRU4REC, li2017neural}, Convolutional Neural Network (CNN)~\cite{tang2018personalized}, Transformer~\cite{SASREC,BERT4REC}, Multilayer Perceptron (MLP)~\cite{FMLPREC}, and Graph Neural Networks (GNN)~\cite{chang2021sequential,wu2019session}, etc.
In addition to employing better neural network architectures, some studies have introduced additional self-supervised signals through contrastive learning~\cite{chen2020simple, he2020momentum} to help the training of recommendation models~\cite{xie2022contrastive,qiu2022contrastive,chen2022intent}.
Unlike these methods, we leverage search history to enhance S\&R tasks.

\paratitle{Personalized Search.} 
The goal of personalized search is to provide relevant items based on the user's searched queries. 
Early works~\cite{ai2019zero,ai2019explainable} only considered the similarity between queries and items, such as QEM~\cite{ai2019zero} and DREM~\cite{ai2019explainable}.
To better understand users' search intent, the personalized search begins to incorporate user profiles and their search histories~\cite{ai2017learning,ai2019zero,bi2020transformer,liu2022category,CoPPS}.
HEM~\cite{ai2017learning} introduces user embeddings. It learns representations of users and items based on reviews through a generative language model.
AEM~\cite{ai2019zero} further incorporates user search histories. It employs an attention mechanism to aggregate the user history.
ZAM~\cite{ai2019zero} builds upon AEM by incorporating a zero vector to control the degree of personalization.
TEM~\cite{bi2020transformer} replaces the attention mechanism in AEM with a transformer.
Unlike existing approaches, we introduce recommendation history to perform both S\&R tasks.

\paratitle{Joint Search and Recommendation.}
In recent years, there has been a trend toward integrating S\&R. These works primarily fall into two categories:
(a)~Search enhanced recommendation~\cite{NRHUB,IV4REC,Query_SeqRec,SESRec,sun2024search}. 
This type of work utilizes search data as supplementary information to enhance the recommendation performance.
IV4Rec~\cite{IV4REC, si2023enhancing} utilizes causal learning, treating user-searched queries as instrumental variables to reconstruct user and item embeddings.
Query-SeqRec~\cite{Query_SeqRec} is a query-aware model which incorporates user queries to model users’ intent.
SESRec~\cite{SESRec} uses contrastive learning to disentangle similar and dissimilar interests between user S\&R behaviors.
(b)~Unified S\&R~\cite{JSR,USER,SRJgraph,xie2023unifiedssr,gong2023unified}. 
This kind of work performs joint learning of S\&R to enhance the model performance in both scenarios.
JSR~\cite{JSR,JSR2} simultaneously trains two models for S\&R using a joint loss function.
USER~\cite{USER} integrates user S\&R behaviors and feeds them into a transformer encoder.
UnifiedSSR~\cite{xie2023unifiedssr} proposes a dual-branch network to encode the product history and query history in parallel.
In this paper, we develop a framework that utilizes a unified set of parameters to handle both S\&R tasks.

\section{Problem Formulation}

Let $\mathcal{U},\mathcal{I},\mathcal{Q}$ denote the sets of users, items, and queries, respectively. Each user $u \in \mathcal{U}$ has a chronologically ordered interaction history $S_u$ that encapsulates all her past S\&R behaviors.
$S_u = \{(x_1,b_1), (x_2,b_2), \dots, (x_N,b_N)\}$, 
where $N$ stands for the number of behaviors exhibited by user $u$. 
The variable $b_t \in \{1,0\}$ signifies the type of the $t$-th behavior, with 1 indicating a recommendation and 0 indicating a search.
Here, $x_t$ denotes the $t$-th behavior: 
\begin{equation*}
x_t=
\begin{cases}
    i_t, & \text{if}~b_t=1~(\text{recommendation}), \\
    \langle q_t, C_{q_t} \rangle, & \text{if}~b_t=0~(\text{search}),
\end{cases}
\end{equation*}
where $i_t \in \mathcal{I}$ represents the $t$-th interacted item, and 
$q_t \in \mathcal{Q}$ is the $t$-th searched query, $C_{q_t} = \{i_1,i_2,\dots,i_{N_{q_t}}\}$ are the $N_{q_t}$ items that $u$ clicked after issuing query $q_t$. 
Please note that $C_{q_t}$ could be an empty set, indicating that $u$ may not click any items ($N_{q_t} = 0$) after issuing $q_t$.

The unified S\&R interactions, denoted as $\mathcal{D}= \mathcal{D}_{R} \bigcup \mathcal{D}_{S}$, contain the recommendation dataset $\mathcal{D}_{R}=\{(u,i_{N+1},q,S_u,y_{u,i_{N+1},q}^{R})_k\}_{k=1}^{|\mathcal{D}_{R}|}$ and the search dataset $\mathcal{D}_{S}=\{(u,i_{N+1},q,S_u,y_{u,i_{N+1},q}^{S})_k\}_{k=1}^{|\mathcal{D}_{S}|}$.
Here, $y_{u,i_{N+1},q}^{R}$ and $y_{u,i_{N+1},q}^{S}$ are user $u$'s preference score for the next item $i_{N+1} \in \mathcal{I}$ after issuing query $q \in \mathcal{Q}$ in recommendation and search scenarios, respectively.
It's worth noting that $q$ is empty for the recommendation scenario.
The objective is to train on dataset $\mathcal{D}$ to find the optimal function $f_{\Theta}$ that predicts $y_{u,i_{N+1},q}^{R}$ and $y_{u,i_{N+1},q}^{S}$ based on the history $S_u$. 
The function can be formulated as follows:
\begin{equation*}
    \left[ \hat{y}^{R}_{u,i_{N+1},q}, ~ \hat{y}^{S}_{u,i_{N+1},q} \right] = f_{\Theta}(u,i_{N+1},q,S_u),
\end{equation*}
where $f_{\Theta}$ denotes our unified model with parameters $\Theta$.  It's important to note that $f_{\Theta}$ have two outputs: $\hat{y}^{R}_{u,i_{N+1},q}$ for recommendation and $\hat{y}^{S}_{u,i_{N+1},q}$ for search.
\section{Our Approach}

\begin{figure*}[t]
    \centering
        \includegraphics[width=0.95\textwidth]{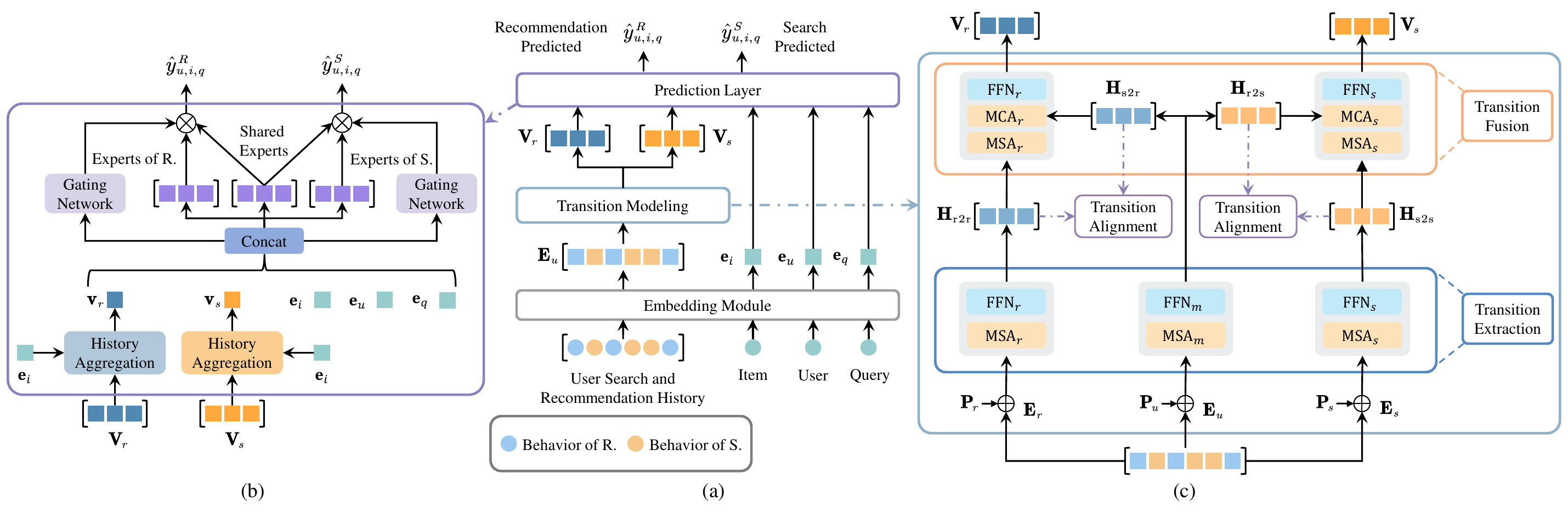}
    \vspace{-8px}
    \caption{
    The architecture of \ourname. (a) \ourname model workflow; (b) Implementation of the prediction layer; (c) The user transition modeling module.
    }
\label{fig:method-model}
\vspace{-0.5cm}
\end{figure*}

This section introduces the proposed method \ourname.
\ourname's overall architecture is shown in Figure~\ref{fig:method-model}.
\ourname first converts the input data into dense representations. 
Then, it employs three distinct transformers using the mask mechanism to separately extract different transitions between user S\&R behaviors. 
Subsequently, contrastive learning is used to align different transitions to learn the relationships between them.
Cross-attentions are also used to fuse these different transitions to obtain the final representations of S\&R histories. 
Following this, an attention mechanism is used to aggregate the historical representations.
Finally, \ourname is jointly trained on S\&R data, enabling its application in both scenarios. MMoE is employed to address the seesaw phenomenon for multi-task training.

\subsection{Embedding Module}
\subsubsection{Embedding Layer}
We have three embedding tables: $\mathbf{M}_U \in \mathbb{R}^{|\mathcal{U}| \times d}$, $\mathbf{M}_I \in \mathbb{R}^{|\mathcal{I}| \times d}$ and $\mathbf{M}_W \in \mathbb{R}^{|\mathcal{W}| \times d}$ for users, items, and the words of all queries. Here, $\mathcal{W}$ is the set composed of all the words included in the queries. $d$ is the embedding dimension. 
Given a user and an item, through the lookup operation, we can get their embeddings: $\mathbf{e}_u \in \mathbb{R}^{d}$ and $\mathbf{e}_i \in \mathbb{R}^{d}$. 
For each query $q$, it contains several words $\{ w_1, w_2, \dots, w_{|q|}\} \subseteq \mathcal{W}$. 
Due to the infrequency of repeated queries in the search data and the typically brief nature of each query, lacking discernible sequential patterns as noted in~\cite{chen2021towards,xie2023unifiedssr}, we derive the query embedding directly by performing mean pooling over the word embeddings: $\mathbf{e}_q = \mathrm{Mean}(\mathbf{e}_{w_1},\mathbf{e}_{w_2}, \dots,\mathbf{e}_{w_{|q|}}) \in \mathbb{R}^{d}$, where $\mathbf{e}_{w_t} \in \mathbb{R}^{d}$ refers to the ID embedding of the $t$-th word.

Regarding the embeddings of behaviors within the user history $S_u$, we directly use the item embeddings as the embeddings for recommendation behaviors. For search behaviors, we combine the query embeddings with the mean pooling of clicked item embeddings to form the embeddings.
The process is detailed below:
\begin{equation}
\label{eq:embedding}
\mathbf{e}_{x_t}=
\begin{cases}
    \mathbf{e}_{i_t}, & \text{if}~b_t=1~(\text{recommendation}), \\
    \mathbf{e}_{q_t}+\mathrm{Mean}(C_{q_t}), & \text{if}~b_t=0~(\text{search}),
\end{cases}
\end{equation}
where $\mathrm{Mean}(C_{q_t})=\mathrm{Mean}(\mathbf{e}_{i_1},\mathbf{e}_{i_2},\dots,\mathbf{e}_{i_{N_{q_t}}})$.
Following this, we obtain the representation $\mathbf{E}_u=[\mathbf{e}_1,\mathbf{e}_2,\ldots,\mathbf{e}_N]^{\intercal} \in \mathbb{R}^{N \times d}$ for the user history $S_u$.

\subsubsection{Query-Item Contrastive Learning}
Search and recommendation behaviors exhibit distinct characteristics.
Most notably, search behavior includes queries, while recommendation does not.
Merging these behaviors directly poses a challenge for the model to capture the transitions between S\&R. 
Hence, we employ contrastive learning to align the representations of queries and items into the same space. 
This enables the model to capture the relationship between S\&R behaviors better. 
Moreover, this approach helps the model comprehend the relevance between queries and items, which is crucial for search performance.

For the $\langle q, C_q \rangle$ pairs specifically, we treat the query $q$ and items in $C_q$ as positive samples, while randomly sampling other queries and items as negative samples. The contrastive loss is computed as~follows: 
\begin{equation}
\label{eq:q_i_cl}
\begin{aligned}
    \mathcal{L}_{\mathrm{Rel}}=
    - & \left[ \sum_{i \in C_q} \mathrm{log}\frac{\mathrm{exp}(\mathrm{sim}(\mathbf{e}_q,\mathbf{e}_i)/\tau_1)}{\sum_{i^{-} \in \mathcal{I}_{\mathrm{neg}}}\mathrm{exp}(\mathrm{sim}(\mathbf{e}_q,\mathbf{e}_{i^{-}})/\tau_1)} \right.\\
    & \left. +~~\sum_{i \in C_q}\mathrm{log}\frac{\mathrm{exp}(\mathrm{sim}(\mathbf{e}_q,\mathbf{e}_i)/\tau_1)}{\sum_{q^{-} \in \mathcal{Q}_\mathrm{neg}}\mathrm{exp}(\mathrm{sim}(\mathbf{e}_{q^{-}},\mathbf{e}_{i})/\tau_1)} \right],
\end{aligned}
\end{equation}
where $\tau_1$ is a learnable temperature coefficient, $\mathcal{I}_\mathrm{neg}$ and $\mathcal{Q}_\mathrm{neg}$ are the set of randomly sampled negative items and queries, respectively. Here, $\mathrm{sim}(\mathbf{a},\mathbf{b})=\mathrm{tanh}(\mathbf{a}\mathbf{W}\mathbf{b}^T)$ computes the similarity between $\mathbf{a} \in \mathbb{R}^d$ and $\mathbf{b} \in \mathbb{R}^d$. The $\mathrm{tanh}$ is the activate function and $\mathbf{W} \in \mathbb{R}^{d \times d}$ maps $\mathbf{a}$ and $\mathbf{b}$ to the representation space where the contrastive loss is calculated.

\subsection{Transition Modeling}
In this section, we introduce the transition modeling module in \ourname, which explicitly models different types of fine-grained user transition behaviors between S\&R. It consists of three components: (1) The \emph{extraction} utilizes transformers equipped with mask mechanism to extract different transitions; (2) The \emph{alignment} employs contrastive learning to align different transitions to learn the relationships between them; (3) The \emph{fusion} uses cross-attention to fuse different transitions to get the overall user representations.

\subsubsection{Transition Extraction}
Firstly, in order to model the transitions of \sts and \rtr, we extract the S\&R sub-histories from $S_u$, 
denoted as $S_s=\{(x_{1},x_{2},\dots,x_{N_s})\}$ and $S_r=\{(x_{1},x_{2},\dots,x_{N_r})\}$.
These sub-histories,
$S_s$ or $S_r$, contain all chronologically ordered search or recommendation behaviors within $S_u$, respectively.
Here, $N_s$ and $N_r$ represent the number of search and recommendation behaviors within the user history $S_u$, and they satisfy the condition $N_s+N_r=N$.
Then, according to Eq.~\eqref{eq:embedding}, we derive their corresponding representations: $\mathbf{E}_s=[\mathbf{e}_{1},\mathbf{e}_{2},\ldots,\mathbf{e}_{N_s}]^{\intercal} \in \mathbb{R}^{N_s \times d}$ and $\mathbf{E}_r=[\mathbf{e}_{1},\mathbf{e}_{2},\ldots,\mathbf{e}_{N_r}]^{\intercal} \in \mathbb{R}^{N_r \times d}$.

To capture the sequential relationships, we further introduce positional embeddings: $\mathbf{P}_s \in \mathbb{R}^{N_s \times d}$, $\mathbf{P}_r \in \mathbb{R}^{N_r \times d}$, and $\mathbf{P}_{u} \in \mathbb{R}^{N \times d}$ for the sequences $S_s$, $S_r$ and $S_u$ respectively.
The final representations of the three sequences are computed as follows:
\begin{equation*}
    \mathbf{\widehat{E}}_s = \mathbf{E}_s + \mathbf{P}_s, \quad
    \mathbf{\widehat{E}}_r = \mathbf{E}_r + \mathbf{P}_r, \quad
    \mathbf{\widehat{E}}_u = \mathbf{E}_u + \mathbf{P}_u.
\end{equation*}

For the transitions of \sts and \rtr, they respectively model the transition within the S\&R sub-histories $S_s$ and $S_r$. 
So we feed representations $\mathbf{\widehat{E}}_s$ and $\mathbf{\widehat{E}}_r$ of $S_s$ and $S_r$ into two transformer~\cite{vaswani2017attention} encoders for their modeling. Each encoder consists of a Multi-head Self-Attention (MSA) layer and a Feed-Forward layer (FFN).
For \sts, MSA takes $\mathbf{\widehat{E}}_s$ as the input, where $\mathbf{Q}=\mathbf{K}=\mathbf{V}=\mathbf{\widehat{E}}_s$.
For \rtr, MSA takes $\mathbf{\widehat{E}}_r$ as its input.
The details are as follows:
\begin{small}
\begin{equation}
\label{eq:MSA_r_s}
\begin{aligned}
\mathbf{H}_{\mathrm{s2s}}=\mathrm{FFN}_s(\mathrm{MSA}_s(\mathbf{\widehat{E}}_s,\mathbf{\widehat{E}}_s,\mathbf{\widehat{E}}_s)),~
\mathbf{H}_{\mathrm{r2r}}=\mathrm{FFN}_r(\mathrm{MSA}_r(\mathbf{\widehat{E}}_r,\mathbf{\widehat{E}}_r,\mathbf{\widehat{E}}_r)), 
\end{aligned}
\end{equation}
\end{small}
where $\mathbf{H}_{\mathrm{s2s}} \in \mathbb{R}^{N_s \times d}$ and $\mathbf{H}_{\mathrm{r2r}} \in \mathbb{R}^{N_r \times d}$ capture the transitions for \sts and \rtr, respectively.

\begin{figure}[t]
    \centering
    \includegraphics[width=0.95\columnwidth]{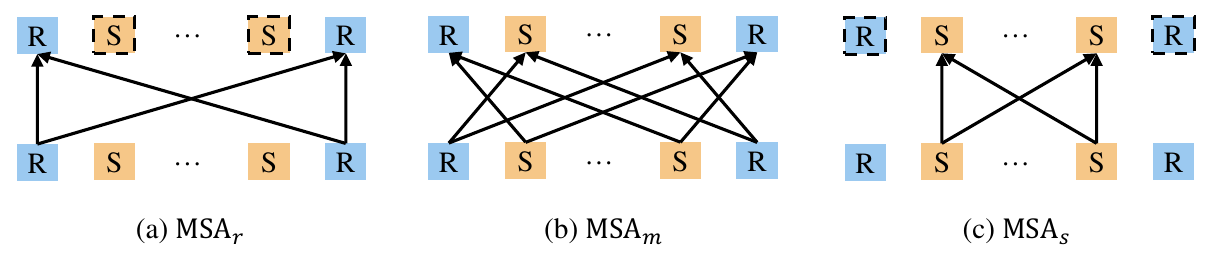}
    \vspace{-5px}
    \caption{
    The attention computation methods differ across various MSA (Multi-head Self-Attention) modules within \ourname. In $\mathrm{MSA}_r$ or $\mathrm{MSA}_s$, attention is specifically computed within recommendation or search behaviors to extract \rtr or \sts, respectively, as detailed in Eq.~\eqref{eq:MSA_r_s}; $\mathrm{MSA}_m$ computes attention exclusively between different behaviors, outlined in Eq.~\eqref{eq:MSA_m}, thereby extracting \rts and \str.
    }
    \label{fig:model-mask}
    \vspace{-0.5cm}
\end{figure}

For the transitions of \rts and \str, they capture the cross-behavior transitions between search and recommendation.
Directly inputting the mixed sequence $S_u$ into a transformer makes it challenging to directly model these transitions because its output would not only include \rts and \str but also \sts and \rtr, causing redundancy with $\mathbf{H}_{\mathrm{s2s}}$ and $\mathbf{H}_{\mathrm{r2r}}$. 
To disentangle different transitions and enable the model to better capture different transitions separately, we introduce a 
mask matrix $\mathbf{M} \in \{0,1\}^{N \times N}$.
The values of $\mathbf{M}$ are as follows:
\begin{equation*}
\mathbf{M}_{ij}=
\begin{cases}
    0, & \text{if}~b_i=b_j, \\
    1, & \text{if}~b_i \neq b_j .
\end{cases}
\end{equation*}
We feed the representations $\mathbf{\widehat{E}}_u$ of $S_u$ and $\mathbf{M}$ into a single transformer encoder. 
Within this encoder, the MSA takes $\mathbf{\widehat{E}}_u$ and $\mathbf{M}$ as inputs, where $\mathbf{Q}=\mathbf{K}=\mathbf{V}=\mathbf{\widehat{E}}_u$:
\begin{equation}
\label{eq:MSA_m}
\mathbf{H}_{m}=\mathrm{FFN}_m(\mathrm{MSA}_m(\mathbf{\widehat{E}}_u,\mathbf{\widehat{E}}_u,\mathbf{\widehat{E}}_u,\mathbf{M})).
\end{equation}
With the incorporation of the mask matrix $\mathbf{M}$ in the input of $\mathrm{MSA}_m$, the computation of the attention in $\mathrm{MSA}_m$ is as follows:
\begin{equation*}
\mathrm{Attention}(\mathbf{\widehat{E}}_u,\mathbf{\widehat{E}}_u,\mathbf{\widehat{E}}_u, \mathbf{M}) = \mathrm{Softmax}\left(\mathbf{\widehat{E}}_u \mathbf{\widehat{E}}_u^{\intercal}/\sqrt{d/h} \odot \mathbf{M} \right) \mathbf{\widehat{E}}_u,
\end{equation*}
where $\odot$ denotes the Hadamard product, and $h$ denotes the number of heads in the $\mathrm{MSA}_m$.
Figure~\ref{fig:model-mask} further demonstrates the varied attention computation methods in different MSA after incorporating masks.
In Eq.~\eqref{eq:MSA_m}, $\mathbf{H}_m \in \mathbb{R}^{N \times d}$ captures transitions specific to \rts and \str, respectively.
The mask matrix $\mathbf{M}$ ensures that attention is computed exclusively among different behaviors within $\mathrm{MSA}_m$, thereby effectively capturing the transitions for \rts and \str.
Finally, we extract the search and recommendation sub-sequences from $\mathbf{H}_m$, yielding $\mathbf{H}_{\mathrm{r2s}} \in \mathbb{R}^{N_s \times d}$ and $\mathbf{H}_{\mathrm{s2r}} \in \mathbb{R}^{N_r \times d}$. These matrices encapsulate the transitions for \rts and \str.

\subsubsection{Transition Alignment.}
\label{sec:cl_before_fusion}
Based on the representations of these four different types of transitions extracted, we further align the varying representations of the same behavior types among users.
For example, $\mathbf{H}_{\mathrm{s2s}}$ and $\mathbf{H}_{\mathrm{r2s}}$ represent the transitions from search and recommendation behaviors in the context of search history, respectively.
Considering that both of them contain information related to the search history, we guide the model to bring their semantics closer.
We utilize contrastive learning to align transitions from the same scenarios with those from different scenarios, enabling the model to learn the correlations between them. This can facilitate a better fusion of various transitions, which is discussed in ~Section~\ref{sec:trans_fusion}.

Specifically, for $\mathbf{H}_{\mathrm{s2s}}$ and $\mathbf{H}_{\mathrm{r2s}}$, they respectively involve transitions from the same and different scenarios in terms of search history. We treat them as positive samples for each other and align them.
Initially, we perform mean pooling on $\mathbf{H}_{\mathrm{s2s}} $ and $\mathbf{H}_{\mathrm{r2s}}$ to obtain $\mathbf{h}_{\mathrm{s2s}} \in \mathbb{R}^{d}$ and $\mathbf{h}_{\mathrm{r2s}} \in \mathbb{R}^{d}$, i.e., $\mathbf{h}_{\mathrm{s2s}}=\mathrm{Mean}(\mathbf{H}_{\mathrm{s2s}})$ and $\mathbf{h}_{\mathrm{r2s}}=\mathrm{Mean}(\mathbf{H}_{\mathrm{r2s}})$.
Then, we treat $\mathbf{h}_{\mathrm{s2s}}$ and $\mathbf{h}_{\mathrm{r2s}}$ as positive samples, while $\mathbf{h}^{-}_{\mathrm{s2s}}$ and $\mathbf{h}^{-}_{\mathrm{r2s}}$ from other users' histories within the same batch as negative samples. Then, the contrastive loss is computed as follows:
\begin{equation}
\begin{aligned}
\mathcal{L}_{\mathrm{Align}}^{S} = - & \left[ \mathrm{log}\frac{\mathrm{exp}(\mathrm{sim}(\mathbf{h}_{\mathrm{s2s}},\mathbf{h}_{\mathrm{r2s}})/\tau_2)}{\sum_{\mathbf{h}^{-}_{\mathrm{r2s}} \in \mathcal{H}_\mathrm{neg}^{\mathrm{r2s}}} \mathrm{exp}(\mathrm{sim}(\mathbf{h}_{\mathrm{s2s}},\mathbf{h}^{-}_{\mathrm{r2s}})/\tau_2)} \right. \\  
& \left. +~~\mathrm{log}\frac{\mathrm{exp}(\mathrm{sim}(\mathbf{h}_{\mathrm{s2s}},\mathbf{h}_{\mathrm{r2s}})/\tau_2)}{\sum_{\mathbf{h}^{-}_{\mathrm{s2s}} \in \mathcal{H}_\mathrm{neg}^{\mathrm{s2s}}} \mathrm{exp}(\mathrm{sim}(\mathbf{h}^{-}_{\mathrm{s2s}},\mathbf{h}_{\mathrm{r2s}})/\tau_2)} \right] ,
\end{aligned}
\end{equation}
where $\tau_2$ is a learnable temperature coefficient, $\mathcal{H}_\mathrm{neg}^{\mathrm{r2s}}$ and $\mathcal{H}_\mathrm{neg}^{\mathrm{s2s}}$ are the set of in-batch negative samples, and the function $\mathrm{sim}(\mathbf{a},\mathbf{b})$ is defined the same as in Eq.~\eqref{eq:q_i_cl}.
Similarly, we obtain the contrastive loss $\mathcal{L}_{\mathrm{Align}}^{R}$ for $\mathbf{H}_{\mathrm{r2r}}$ and $\mathbf{H}_{\mathrm{s2r}}$.
The total contrastive loss for transition alignment is:
\begin{equation}
\label{eq:cl_fusion}
    \mathcal{L}_{\mathrm{Align}} = \mathcal{L}_{\mathrm{Align}}^{S} + \mathcal{L}_{\mathrm{Align}}^{R}.
\end{equation}

\subsubsection{Transition Fusion}
\label{sec:trans_fusion}
After extracting and learning the relationships between different transitions, we fuse them to get the overall user representation.
Specifically, we fuse $\mathbf{H}_{\mathrm{s2s}}$ with $\mathbf{H}_{\mathrm{r2s}}$ as well as $\mathbf{H}_{\mathrm{r2r}}$ with $\mathbf{H}_{\mathrm{s2r}}$ to derive the final representations for S\&R histories.  Drawing inspiration from prior work on multi-modal fusion~\cite{li2021align,dou2022empirical}, we utilize Multi-head Cross-Attention (MCA) for this fusion. 
For search history, MCA takes $\mathbf{H}_{\mathrm{s2s}}$ with $\mathbf{H}_{\mathrm{r2s}}$ as the inputs for fusion, where $\mathbf{Q}=\mathbf{H}_{\mathrm{r2s}}, \mathbf{K}=\mathbf{V}=\mathbf{H}_{\mathrm{s2s}}$.
While, for recommendation history, MCA takes $\mathbf{H}_{\mathrm{r2r}}$ with $\mathbf{H}_{\mathrm{s2r}}$ as the inputs.
The specific process is as~follows:
\begin{small}
\begin{equation}
\label{eq:trans_fusion}
\begin{aligned}
\mathbf{F}_s&= \mathrm{MSA}_s(\mathbf{H}_{\mathrm{s2s}},\mathbf{H}_{\mathrm{s2s}},\mathbf{H}_{\mathrm{s2s}}), ~
\mathbf{V}_s=\mathrm{FFN}_s(\mathrm{MCA}_s(\mathbf{H}_{\mathrm{r2s}},\mathbf{F}_{s},\mathbf{F}_s)), \\
\mathbf{F}_r&= \mathrm{MSA}_r(\mathbf{H}_{\mathrm{r2r}},\mathbf{H}_{\mathrm{r2r}},\mathbf{H}_{\mathrm{r2r}}), ~
\mathbf{V}_r=\mathrm{FFN}_r(\mathrm{MCA}_r(\mathbf{H}_{\mathrm{s2r}},\mathbf{F}_{r},\mathbf{F}_r)),
\end{aligned}
\end{equation}
\end{small}
where $\mathbf{V}_s \in \mathbb{R}^{N_s \times d}$ and $\mathbf{V}_r \in \mathbb{R}^{N_r \times d}$ are the final representations for S\&R histories, respectively. The two representations constitute the overall user representation.

\subsection{Model Prediction and Training}

\subsubsection{History Aggregation}
\label{sec:TransNext}
After obtaining representations $\mathbf{V}_s$ and $\mathbf{V}_r$ in Eq.~\eqref{eq:trans_fusion} for S\&R histories, when predicting whether the next item $i$ will be clicked, we aggregate the history using attention mechanism to extract user interest from the transitions concerning the target item $i$.
The calculation of the attention is as follows:
\begin{equation}
\label{eq:target_attn}
    \mathbf{v}_s = \mathbf{V}_s^{\intercal} \mathrm{Softmax}(\mathbf{V}_s \mathbf{W}_s \mathbf{e}_i), \quad
    \mathbf{v}_r = \mathbf{V}_r^{\intercal} \mathrm{Softmax}(\mathbf{V}_r \mathbf{W}_r \mathbf{e}_i)
\end{equation}
where $\mathbf{W}_s \in \mathbb{R}^{d \times d}$ and $\mathbf{W}_r \in \mathbb{R}^{d \times d}$ are learnable parameters, $\mathbf{v}_s \in \mathbb{R}^d$ and $\mathbf{v}_r \in \mathbb{R}^d$ are the aggregated representations for S\&R histories.
The attention mechanism in Eq.~\eqref{eq:target_attn} computes the similarities between the target item and historical behaviors; the higher the similarities, the greater the influence of historical behaviors on the transition to the target item.

\subsubsection{Prediction}
\label{sec:predict}
\ourname adopts a unified model for S\&R tasks.
Building on the previous modules' modeling of user transition behaviors, we can predict two tasks separately.
In the field of information retrieval, a multi-task framework~\cite{ma2018modeling,tang2020progressive,wang2023multi} is commonly used to predict different tasks.
Considering that direct making predictions faces challenges due to the differences in data distributions between S\&R,
we employ a multi-task approach to separately predict the two tasks, incorporating behavior-shared and behavior-specific modules.

\ourname employs a single model to predict both S\&R tasks.
We introduce a learnable query $q_{\phi} \in \mathbb{R}^d$ for the recommendation scenario to unify the input for predictions between S\&R.
Firstly, we concatenate representations of users, items, queries, and the S\&R histories to obtain a shared bottom representation, denoted as $\mathbf{X}_b=\mathrm{Concat}(\mathbf{e}_u,\mathbf{e}_i,\mathbf{e}_q,\mathbf{v}_s,\mathbf{v}_r)$, where $\mathbf{X}_b \in \mathbb{R}^{d_b} $.
Here $d_b = 5\times d$.
Next, to balance the relationship between different tasks, we exploit the Multi-gated Mixture of Experts (MMoE)~\cite{ma2018modeling,tang2020progressive} technique.
We denote the shared and specific expert sets for S\&R as $\mathcal{O}_m$, $\mathcal{O}_s$, and $\mathcal{O}_r$, where the number of experts for each is represented by $n_m$, $n_s$, and $n_r$, respectively.
The output of the MMoE module consists of two scores, corresponding to S\&R, calculated as follows:
\begin{equation}
\label{eq:predict}
\begin{aligned}
    \hat{y}_{u,i,q}^{S} &= \sum_{o \in \mathcal{O}_s}g_o^s(\mathbf{X}_b) o(\mathbf{X}_b) + \sum_{o \in \mathcal{O}_m}g_o^s(\mathbf{X}_b) o(\mathbf{X}_b), \\
    \hat{y}_{u,i,q}^{R} &= \sum_{o \in \mathcal{O}_r}g_o^r(\mathbf{X}_b) o(\mathbf{X}_b) + \sum_{o \in \mathcal{O}_m}g_o^r(\mathbf{X}_b) o(\mathbf{X}_b), 
\end{aligned}
\end{equation}
where $g^s(\mathbf{X}_b) = \mathrm{Softmax}(\mathbf{W}^s \mathbf{X}_b)$ and $g^r(\mathbf{X}_b) = \mathrm{Softmax}(\mathbf{W}^r \mathbf{X}_b)$ are the gating networks for S\&R, $\mathbf{W}^s \in \mathbb{R}^{(n_s + n_m) \times d_b}$ and $\mathbf{W}^r \in \mathbb{R}^{(n_r + n_m) \times d_b}$ are used for linear transformation.
Here $o(\cdot)$ denotes the expert networks, Multi-Layer Perceptrons (MLPs) in our implementation. 
$g_o^s$ and $g_o^r$ are the weights of expert $o(\cdot)$ generated by the gating network for S\&R.

\begin{table}[t]
    \small
    \centering
    \caption{
    Comparison of different joint S\&R methods. 
    "Explicit" denotes whether the model explicitly considers the differences between different transitions. "One Model" refers to whether a unified set of parameters is used to simultaneously serve both search and recommendation scenarios.}
    \vspace{-5px}
    \label{tab:comparison}
    \resizebox{.95\columnwidth}{!}{
    \begin{tabular}
    {l
     ccccc
     cc
     c
    }
    \toprule
    \multicolumn{1}{l}{\multirow{2}{*}{Methods}} & 
    \multicolumn{5}{c}{Transition Modeling} &
    \multicolumn{2}{c}{Task} & 
    \multicolumn{1}{c}{\multirow{2}{*}{One Model}} \\
    \cmidrule(l){2-6} \cmidrule(l){7-8} 
    \multicolumn{1}{c}{}  
    & \rtr & \sts & \rts & \str & Explicit
    & R. & S.
    \\
    \midrule
    JSR~\cite{JSR} &\textcolor{purple}{\XSolidBrush} &\textcolor{purple}{\XSolidBrush} &\textcolor{purple}{\XSolidBrush} &\textcolor{purple}{\XSolidBrush} &\textcolor{purple}{\XSolidBrush} &\textcolor{teal}{\CheckmarkBold} &\textcolor{teal}{\CheckmarkBold} &\textcolor{purple}{\XSolidBrush} \\    
    SESRec~\cite{SESRec} &\textcolor{teal}{\CheckmarkBold} &\textcolor{teal}{\CheckmarkBold} &\textcolor{purple}{\XSolidBrush} &\textcolor{purple}{\XSolidBrush} &\textcolor{purple}{\XSolidBrush} &\textcolor{teal}{\CheckmarkBold} &\textcolor{purple}{\XSolidBrush} &\textcolor{teal}{\CheckmarkBold} \\
    UnifiedSSR~\cite{xie2023unifiedssr} &\textcolor{teal}{\CheckmarkBold} &\textcolor{teal}{\CheckmarkBold} &\textcolor{purple}{\XSolidBrush} &\textcolor{purple}{\XSolidBrush} &\textcolor{purple}{\XSolidBrush} &\textcolor{teal}{\CheckmarkBold} &\textcolor{teal}{\CheckmarkBold} &\textcolor{purple}{\XSolidBrush} \\
    USER~\cite{USER} &\textcolor{teal}{\CheckmarkBold} &\textcolor{teal}{\CheckmarkBold} &\textcolor{teal}{\CheckmarkBold} &\textcolor{teal}{\CheckmarkBold} &\textcolor{purple}{\XSolidBrush} &\textcolor{teal}{\CheckmarkBold} &\textcolor{teal}{\CheckmarkBold} &\textcolor{purple}{\XSolidBrush} \\
    \ourname (ours) &\textcolor{teal}{\CheckmarkBold} &\textcolor{teal}{\CheckmarkBold} &\textcolor{teal}{\CheckmarkBold} &\textcolor{teal}{\CheckmarkBold} &\textcolor{teal}{\CheckmarkBold} &\textcolor{teal}{\CheckmarkBold} &\textcolor{teal}{\CheckmarkBold} &\textcolor{teal}{\CheckmarkBold} \\
    \bottomrule
    \end{tabular}
    }
    \vspace{-0.5cm}
\end{table}

\subsubsection{Training}
Following previous works~\cite{DIN,DIEN,SESRec}, 
we utilize binary cross-entropy loss to optimize our model: 
\begin{small}
\begin{equation}
\label{eq:loss_click}
\begin{aligned}
    \mathcal{L}_{\mathrm{Click}}^{S} &= -\frac{1}{|\mathcal{D}_S|} \sum_{(u,i,q) \in \mathcal{D}_S} y_{u,i,q}^{S}\mathrm{log}(\hat{y}_{u,i,q}^{S}) + (1-y_{u,i,q}^{S})\mathrm{log}(1-\hat{y}_{u,i,q}^{S}),  \\
    \mathcal{L}_{\mathrm{Click}}^{R} &= -\frac{1}{|\mathcal{D}_R|} \sum_{(u,i,q) \in \mathcal{D}_R} y_{u,i,q}^{R}\mathrm{log}(\hat{y}_{u,i,q}^{R}) + (1-y_{u,i,q}^{R})\mathrm{log}(1-\hat{y}_{u,i,q}^{R}),
\end{aligned}
\end{equation}
\end{small}
where $\mathcal{L}_{\mathrm{Click}}^{S}$ and $\mathcal{L}_{\mathrm{Click}}^{R}$ are the click loss for S\&R respectively.
The total loss for S\&R includes the click loss in Eq.~\eqref{eq:loss_click} and the contrastive loss in Eq.~\eqref{eq:q_i_cl} and Eq.~\eqref{eq:cl_fusion}:
\begin{equation}
\label{eq:loss_rec}
\begin{aligned}
\mathcal{L}_{S} &= \mathcal{L}_{\mathrm{Click}}^{S} + \alpha \mathcal{L}_{\mathrm{Rel}} + \beta \mathcal{L}_{\mathrm{Align}}, \\
\mathcal{L}_{R} &= \mathcal{L}_{\mathrm{Click}}^{R} + \alpha \mathcal{L}_{\mathrm{Rel}} + \beta \mathcal{L}_{\mathrm{Align}},
\end{aligned}
\end{equation}
where $\alpha$ and $\beta$ are the hyper-parameters for contrastive loss. 
Finally, the overall loss for joint training of S\&R is:
\begin{equation}
\label{eq:loss_total}
    \mathcal{L}_{\mathrm{Total}}= \mathcal{L}_{R} + \gamma \mathcal{L}_{S} + \lambda ||\Theta||_2,
\end{equation}
where $\gamma$ controls the trade-off between S\&R, $\Theta$ denotes the parameters of \ourname, $\lambda$ is a hyper-parameter which controls the $L_2$~regularization.

\subsection{Discussion}

Table~\ref{tab:comparison} summarized the ability of \ourname and existing joint S\&R models in terms of modeling user S\&R histories in user modeling.
JSR~\cite{JSR,JSR2} neglects the transitions between users' different behaviors.
SESRec~\cite{SESRec} and UnifiedSSR~\cite{xie2023unifiedssr} separately input the S\&R histories into two encoders, modeling the transitions of \sts and \rtr.
USER~\cite{USER} mixes the user S\&R histories into one sequence and implicitly captures the transitions through a single encoder. 
In USER, different transitions are modeled in the same manner.
\ourname explicitly models fine-grained user transitions in different ways, which considers the differences between various transitions.

Furthermore, JSR employs two sets of parameters that are jointly trained by sharing the item set. SESRec utilizes search as supplementary information to enhance recommendations. USER and UnifiedSSR derive two parameter sets for S\&R tasks via pre-training and fine-tuning. 
In this paper, we leverage MMoE to achieve optimal performance in both S\&R tasks using a single unified set of parameters, reducing parameter redundancy. In summary, \ourname is more efficient and effective.
\section{Experiments}
We conducted experiments on two public datasets to evaluate the performance of \ourname. 
The source code is available \footnote{\url{https://github.com/TengShi-RUC/UniSAR}}.

\begin{table}[t]
    \caption{Statistics of the datasets used in this paper.}
    \vspace{-8px}
    \center
     \resizebox{.95\columnwidth}{!}{
        \begin{tabular}{cccccc}
        \toprule
        Dataset & \#Users & \#Items & \#Queries & \#Action-S &\#Action-R  \\
        \midrule
        KuaiSAR & 25,877 & 6,890,707 & 453,667 & 5,059,169 & 14,605,716 \\
        Amazon  & 68,223 & 61,934 & 4,298 & 934,664 & 989,618 \\
        \bottomrule
        \end{tabular}}
    \label{tab:dataStatistics}   
    \vspace{-0.5cm}
\end{table}

\subsection{Experimental Setup}

\subsubsection{Dataset} 
The experiments were conducted on the following publicly available datasets.
The statistics of these datasets are shown in Table~\ref{tab:dataStatistics}.

\textbf{KuaiSAR}\footnote{\href{https://kuaisar.github.io/}{https://kuaisar.github.io/}}~\cite{Sun2023KuaiSAR}: 
The dataset offers authentic user S\&R behaviors.
As for pre-processing, we only kept users who exhibited both S\&R behaviors and filtered data of the last two days due to their sparse search behaviors. 
Within the remaining data, following~\cite{SASREC,FMLPREC}, we filtered items and users with fewer than five interaction records.
Following~\cite{zhao2020revisiting}, we used the data from the last day for testing, the second last day for validation, and the rest for training.

\textbf{Amazon}\footnote{\href{http://jmcauley.ucsd.edu/data/amazon/}{http://jmcauley.ucsd.edu/data/amazon/}}~\cite{amazon_dataset, amazon_dataset2}: 
We also adopted a widely accepted semi-synthetic dataset.  
Following previous works~\cite{ai2017learning,ai2019zero,SESRec,JSR}, we generated synthetic search behaviors for this recommendation dataset. We adopt the ``Kindle Store'' subset of the five-core Amazon dataset that covers data in which all users and items have at least five interactions. 
We strictly adhered to the established practices, as mentioned in~\cite{SESRec}, for data construction and the leave-one-out strategy to split the dataset.

\begin{table*}[h!]
\small
\centering
\caption{Comparisons of the overall \emph{recommendation} performance of different methods on both datasets.
The best and the second-best methods are highlighted in bold and underlined fonts, respectively.
* indicates that improvements over the second-best methods are statistically significant ($t$-test, $p$-value$<0.01$).
}
\vspace{-3px}
\label{tab:rec_result}
\resizebox{.98\textwidth}{!}{
\begin{tabular}{llcccccccccccc}
\toprule
\multicolumn{1}{l}{\multirow{2}{*}{Datasets}} & 
\multicolumn{1}{l}{\multirow{2}{*}{Metric}} & 
\multicolumn{5}{c}{\emph{Sequential Recommendation}} & 
\multicolumn{7}{c}{\emph{Joint Search and Recommendation}} \\ 
\cmidrule(l){3-7} \cmidrule(l){8-14}
\multicolumn{1}{c}{}  & \multicolumn{1}{c}{} 
& DIN & GRU4Rec & SASRec & BERT4Rec & FMLP-Rec & NRHUB & Query-SeqRec &SESRec & JSR &USER  &UnifiedSSR & \textbf{\ourname}  \\ 
\midrule
\multirow{7} * {KuaiSAR}
&HR@1 &0.1629 &0.1097 &0.1249 &0.1061 &0.1370 &0.1243 &0.1166 &\underline{0.1827} &0.1754 &0.1489 &0.1225 &\textbf{0.1990}* \\
&HR@5 &0.4509 &0.3764 &0.4065 &0.3699 &0.4292 &0.3862 &0.3920 &\underline{0.4956} &0.4791 &0.4086 &0.3981 &\textbf{0.5169}* \\
&HR@10 &0.6179 &0.5788 &0.6007 &0.5885 &0.6159 &0.5610 &0.5890 &\underline{0.6643} &0.6453 &0.5627 &0.5939 &\textbf{0.6792}* \\
&NDCG@5 &0.3104 &0.2435 &0.2671 &0.2381 &0.2851 &0.2572 &0.2552 &\underline{0.3432} &0.3315 &0.2820 &0.2617 &\textbf{0.3632}* \\
&NDCG@10 &0.3643 &0.3087 &0.3298 &0.3083 &0.3453 &0.3136 &0.3186 &\underline{0.3978} &0.3853 &0.3318 &0.3249 &\textbf{0.4158}* \\
\hline
\multirow{7} * {\makecell[l]{Amazon \\ (Kindle Store)}}
&HR@1 &0.2159 &0.1725 &0.2059 &0.2481 &0.1991 &0.1889 &0.2186 &\underline{0.2726} &0.2346 &0.2361 &0.2013 &\textbf{0.3010}* \\
&HR@5 &0.5170 &0.4949 &0.5295 &0.5311 &0.5356 &0.4988 &0.5401 &\underline{0.5623} &0.5467 &0.5441 &0.5196 &\textbf{0.5874}* \\
&HR@10 &0.6525 &0.6548 &0.6772 &0.6658 &\underline{0.6879} &0.6503 &0.6860 &0.6864 &0.6779 &0.6854 &0.6707 &\textbf{0.7020}* \\
&NDCG@5 &0.3726 &0.3388 &0.3747 &0.3954 &0.3739 &0.3487 &0.3859 &\underline{0.4245} &0.3970 &0.3964 &0.3662 &\textbf{0.4513}* \\
&NDCG@10 &0.4165 &0.3907 &0.4225 &0.4390 &0.4232 &0.3977 &0.4333 &\underline{0.4648} &0.4396 &0.4422 &0.4151 &\textbf{0.4885}* \\
\bottomrule
\end{tabular} 
}
\vspace{-0.3cm}
\end{table*}

\subsubsection{Evaluation Metrics}
Following previous works~\cite{BERT4REC, FMLPREC, SESRec}, we utilize the ranking metrics, \textit{Hit Ratio} (HR) and \textit{Normalized Discounted Cumulative Gain} (NDCG).
We report HR at positions $\{1,5,10\}$ and NDCG at positions $\{5,10\}$.
Following the common strategy~\cite{SASREC, FMLPREC, SESRec}, we pair the ground-truth item with 99 randomly sampled negative items that the user has not interacted~with.

\subsubsection{Baselines}
In this work, we compare \ourname with the following state-of-the-art methods.

Firstly, we compare our methods with the following \textbf{sequential recommendation models} that do not utilize search data, including 
(1) \textbf{DIN}~\cite{DIN} uses a local activation unit to adaptively learn the representation of user interests;
(2) \textbf{GRU4Rec}~\cite{GRU4REC} applies GRUs to model users' interaction history.
(3) \textbf{SASRec}~\cite{SASREC} is a unidirectional Transformer-based model; 
(4) \textbf{BERT4Rec}~\cite{BERT4REC} uses a cloze objective loss for sequential recommendation by the bidirectional Transformer;
and (5) \textbf{FMLPRec}~\cite{FMLPREC} is an all-MLP model with learnable filters;

Secondly, we compare our methods with the following \textbf{personalized search models} that do not utilize recommendation data, including
(6) \textbf{QEM}~\cite{ai2019zero} only considers the matching scores between items and queries;
(7) \textbf{HEM}~\cite{ai2017learning} is a latent vector-based personalized model;
(8) \textbf{AEM}~\cite{ai2019zero} is an attention-based personalized model that aggregates the user's historical interacted items with the current query;
(9) \textbf{ZAM}~\cite{ai2019zero} improves AEM by concatenating a zero vector to the item list.
(10) \textbf{TEM}~\cite{bi2020transformer} upgrades the attention layer in AEM with transformer encoder;
and (11) \textbf{CoPPS}~\cite{CoPPS} uses contrastive learning techniques;

Finally, we compare our methods with the following two classes of \textbf{joint S\&R models}:
(a) \textbf{search enhanced recommendation}:
(12) \textbf{NRHUB}~\cite{NRHUB} is a news recommendation model leveraging heterogeneous user behaviors;
(13) \textbf{Query-SeqRec}~\cite{Query_SeqRec} is a query-aware sequential model which incorporates queries into user behaviors using transformers;
(14) \textbf{SESRec}~\cite{SESRec} learns disentangled search representation for recommendation using contrastive learning;
(b) \textbf{unified S\&R}:
(15) \textbf{JSR}~\cite{JSR} is a general framework which optimizes a joint loss. Following~\cite{IV4REC}, the recommendation model in JSR is implemented using DIN in our experiment;
(16) \textbf{USER}~\cite{USER} integrates the user’s behaviors in S\&R into a heterogeneous behavior sequence.
(17) \textbf{UnifiedSSR}~\cite{xie2023unifiedssr} jointly learns the user behavior history in both S\&R scenarios.

\subsubsection{Implementation Details}
All hyper-parameters for the baselines are searched according to the settings in the original papers. 
The embedding dimension $d$ is set to 64 for both the KuaiSAR and Amazon datasets.
For both datasets, the maximum length for S\&R histories is set to 30. 
The numbers of shared and specific experts, $n_m$, $n_s$, and $n_r$ in Section~\ref{sec:predict}, are both set to 4.
The temperature parameters, $\tau_1$ in Eq.~\eqref{eq:q_i_cl} and $\tau_2$ in Eq.~\eqref{eq:cl_fusion}, are tuned among $[0.1:+0.1:1.0]$.
The weight $\alpha$ and $\beta$ in Eq.~\eqref{eq:loss_rec} are tuned among $\{1e\text{-}1, 1e\text{-}2, 1e\text{-}3, 1e\text{-}4, 1e\text{-}5\}$.
The weight $\gamma$ in Eq.~\eqref{eq:loss_total} is tuned among $[0.01,1.0]$.
The batch size is set as 1024.
We train all the models with 100 epochs and adopt early-stopping to avoid over-fitting.
Adam \cite{kingma2014adam} is used to conduct the optimization.
The learning rate is tuned among $\{1e\text{-}3,1e\text{-}4,1e\text{-}5\}$, and $\lambda$ in Eq.~\eqref{eq:loss_total} is searched from $\{1e\text{-}4,1e\text{-}5,1e\text{-}6\}$.

\subsection{Experimental Results}
Tables~\ref{tab:rec_result} and~\ref{tab:src_result} report the results on two datasets of S\&R tasks, respectively.
From the experimental results, we can observe that:

\noindent\textbf{$\bullet $}~Firstly, compared to existing joint S\&R models, \ourname has achieved optimal results on all datasets.
Overall, \ourname outperforms the SOTA models, passing the significance test ($p$-value $< 0.01$) in most cases.
This demonstrates the effectiveness of modeling various user transitions and introducing MMoE to facilitate joint training.

\noindent\textbf{$\bullet $}~Secondly, compared to existing models for sequential recommendation or personalized search, \ourname has consistently yielded superior results. This demonstrates that the integration of S\&R in \ourname benefits both tasks. 
Furthermore, we can observe that not all models combining S\&R yield better results than typical sequential recommendation or personalized search models. For instance, NRHUB and Query-SeqRec don't exhibit this trend. This indicates that simply blending S\&R data doesn't suffice. Instead, a fine-grained consideration of transitions between user S\&R behaviors is necessary.

\noindent\textbf{$\bullet $}~Finally, on KuaiSAR data, the model jointly trained for S\&R significantly outperforms the personalized search model that only considers search history. This is because, as indicated by the statistics in Table~\ref{tab:dataStatistics}, 
the KuaiSAR search data is extremely sparse.
Joint training allows the utilization of recommendation information to alleviate the sparsity of search data, thus leading to a substantial improvement in search performance.

\begin{table*}[h!]
\small
\centering
\caption{Comparisons of the overall \emph{search} performance of different methods on both datasets.
The best and the second-best methods are highlighted in bold and underlined fonts, respectively.
* indicates that improvements over the second-best methods are statistically significant ($t$-test, $p$-value$<0.01$).}
    \vspace{-3px}
    \label{tab:src_result}
 \resizebox{.85\textwidth}{!}{
 \begin{tabular}
 {llccccccccccc}
    \toprule
    \multicolumn{1}{l }{\multirow{2}{*}{Datasets}} & 
    \multicolumn{1}{l }{\multirow{2}{*}{Metric}} & 
 \multicolumn{6}{c}{\emph{Personalized Search}} & 
 \multicolumn{4}{c}{\emph{Joint Search and Recommendation}}  	\\ 
 \cmidrule(l){3-8} \cmidrule(l){9-12}
    \multicolumn{1}{c}{}  & \multicolumn{1}{c}{}         & 
 QEM & HEM & AEM & ZAM & TEM & CoPPS & 
 JSR &USER &UnifiedSSR & \textbf{\ourname} \\ \midrule
\multirow{7} * {KuaiSAR}
&HR@1 &0.2944 &0.3337 &0.2703 &0.2815 &0.3045 &0.3117 &0.4543 &\underline{0.4628} &0.4389 &\textbf{0.5282}* \\
&HR@5 &0.6020 &0.6505 &0.5956 &0.6117 &0.6502 &0.6616 &0.7162 &0.7304 &\underline{0.7377} &\textbf{0.7476}* \\
&HR@10 &0.7182 &0.7653 &0.7182 &0.7344 &0.7632 &0.7707 &0.7961 &0.8149 &\underline{0.8320} &\textbf{0.8369}* \\
&NDCG@5 &0.4575 &0.5029 &0.4415 &0.4560 &0.4887 &0.4977 &0.5962 &\underline{0.6069} &0.5991 &\textbf{0.6417}* \\
&NDCG@10 &0.4953 &0.5400 &0.4812 &0.4959 &0.5254 &0.5331 &0.6221 &\underline{0.6342} &0.6297 &\textbf{0.6708}* \\
\hline
\multirow{7} * {\makecell[l]{Amazon \\ (Kindle Store)}}
&HR@1 &0.2772 &0.2497 &0.2916 &0.2954 &0.4090 &0.4052 &0.3176 &\underline{0.4123} &0.3663 &\textbf{0.5343}* \\
&HR@5 &0.7100 &0.6778 &0.7095 &0.7109 &\underline{0.8185} &0.8169 &0.7038 &0.7631 &0.7744 &\textbf{0.8190} \\
&HR@10 &0.8186 &0.8267 &0.8443 &0.8468 &\textbf{0.9051} &\underline{0.9051} &0.8225 &0.8697 &0.8812 &0.8977 \\
&NDCG@5 &0.5066 &0.4736 &0.5114 &0.5147 &\underline{0.6303} &0.6281 &0.5173 &0.6000 &0.5847 &\textbf{0.6875}* \\
&NDCG@10 &0.5422 &0.5221 &0.5554 &0.5590 &\underline{0.6587} &0.6570 &0.5563 &0.6348 &0.6196 &\textbf{0.7132}* \\
\bottomrule
    \end{tabular}
 }
 \vspace{-0.3cm}
\end{table*}

\begin{table}[t!]
    \small
    \caption{Ablation study on the KuaiSAR dataset. "w/o" stands for "without", indicating that the corresponding module in \ourname is removed.}
    \vspace{-5px}
    \label{tab:ablation_result}
    \renewcommand{\arraystretch}{1.2}
 \resizebox{.95\columnwidth}{!}{
 \begin{tabular}
 {lcccc}
    \toprule
    \multicolumn{1}{l }{\multirow{2}{*}{Model}} & 
 \multicolumn{2}{c}{\emph{Recommendation}} & 
 \multicolumn{2}{c}{\emph{Search}}  	\\ 
 \cmidrule(l){2-3} \cmidrule(l){4-5}
    \multicolumn{1}{c}{}   
&NDCG@5 &NDCG@10
&NDCG@5 &NDCG@10  \\ 
 \midrule
\textbf{\ourname} &\textbf{0.3632} &\textbf{0.4158} &\textbf{0.6417} &\textbf{0.6708} \\
\hline
w/o \rtr &0.3258 &0.3798 &0.6007 &0.6319  \\
w/o \rts &0.3606 &0.4136 &0.6376 &0.6690  \\
w/o \str &0.3512 &0.4036 &0.6292 &0.6584  \\
w/o \sts &0.3399 &0.3941 &0.6404 &0.6706  \\
w/o mask $\mathbf{M}$ &0.3503 &0.4024 &0.6217 &0.6507 \\
\hdashline
w/o $\mathcal{L}_{\mathrm{Align}}$ &0.3458 &0.3956 &0.6069 &0.6381 \\
w/o $\mathcal{L}_{\mathrm{Rel}}$ &0.3472 &0.3988 &0.6177 &0.6465  \\
\hdashline
w/o $\mathrm{MCA}_r$ &0.3384 &0.3924 &0.6094 &0.6397 \\
w/o $\mathrm{MCA}_s$ &0.3561 &0.4091 &0.6257 &0.6545 \\
\hdashline
w/o MMoE &0.3132 &0.3692 &0.6143 &0.6438 \\
w/o Joint Training
&0.3569 &0.4105 &0.5850 &0.6167 \\
\bottomrule
\end{tabular}
 } 
 \vspace{-0.5cm}
\end{table}

\subsection{Ablation Study}
\label{sec: ablation study}

We conducted an ablation study on KuaiSAR datasets to investigate the effectiveness of different modules in UniSAR.

\textbf{(1) Transition Extraction: } UniSAR extracts four types of transitions from S\&R behaviors, and we removed each transition individually to verify its role.
After removing a transition, we remove its corresponding alignment loss and modify some module inputs to keep the architecture's consistency. For instance, the variant `w/o \rtr' removes $\mathcal{L}_{\mathrm{Align}}^{R}$ in Eq.~\eqref{eq:cl_fusion} and replaces 
$\mathbf{H}_{\mathrm{r2r}}$ with $\mathbf{H}_{\mathrm{s2r}}$ in Eq.~\eqref{eq:trans_fusion}. 
The first four rows in~\autoref{tab:ablation_result} show that removing any transition leads to decreased performances, verifying that each transition is beneficial for both tasks.

To further validate the effectiveness of our design in extracting two types of cross-behavior transitions (i.e., \rts \& \str), we removed the mask matrix $\mathbf{M}$ in Eq.~\eqref{eq:MSA_m}.
The fifth row's results in~\autoref{tab:ablation_result} confirm the effectiveness of separate and fine-grained modeling cross-behavior transitions and the other two types of transitions.

\textbf{(2) Transition Alignment: }
We investigated the impact of transition alignment in Eq.~\eqref{eq:cl_fusion} by removing the alignment loss $\mathcal{L}_{\mathrm{Align}}$.
We observed that removing $\mathcal{L}_{\mathrm{Align}}$ would lead to a decrease in the performance of both S\&R. 
This indicates that $\mathcal{L}_{\mathrm{Align}}$ can assist the model in learning the correlations between different transitions,  leading to an effective fusion of transitions.

Additionally, we investigated the impact of $\mathcal{L}_{\mathrm{Rel}}$ in Eq.~\eqref{eq:q_i_cl}. 
Removing $\mathcal{L}_{\mathrm{Rel}}$ also leads to a decrease in S\&R performance.
This is because $\mathcal{L}_{\mathrm{Rel}}$ helps the model learn the relevance between different queries and items, which forms the basis for the model's semantic alignment of different transitions.

\textbf{(3) Transition Fusion: }
We explore the effects of cross-attention in Section~\ref{sec:trans_fusion}, which are used for transition fusion to get the overall user representations.
The results are shown in the lower part of Table~\ref{tab:ablation_result}.
`w/o $\mathrm{MCA}_r$' indicates that we replace this module with an element-wise addition of $\mathbf{H}_{\mathrm{s2s}}$ with $\mathbf{H}_{\mathrm{r2s}}$.
Similar operations are done for `w/o $\mathrm{MCA}_s$'.
From the results, we observed that the model's performance declined on both tasks, demonstrating the rationality of using cross-attention for fusion.

\textbf{(4) Multi-task Training: }
We explore the impact of MMoE and the joint training of two tasks. 
The results are presented in the bottom part of~\autoref{tab:ablation_result}.
`w/o MMoE' means that we directly remove the MMoE module, and employ two separate MLPs to predict S\&R tasks, respectively. Two tasks share most model parameters except these two MLPs.
`w/o Joint Training' denotes the separate training of S\&R, resulting in two sets of model parameters dedicated to each task. Since each set of parameters handles a single task, MMoE is no longer necessary and hence removed.
The performance of these two variants was significantly lower than UniSAR, indicating that removing MMoE or removing joint training adversely affects the model performance.

Further, by comparing these two variants, we observed a seesaw phenomenon in model performance when training on both tasks.
`w/o MMoE' leads to a significant improvement in search and a large decline in recommendation compared to `w/o Joint Training'. 
This means that joint training improves the performance of one task but reduces that of the other.
This is because, in KuaiSAR, the volume of interactions in recommendation data is much greater than that in search data.
There is a trade-off between two tasks. Incorporating MMoE helps us overcome this issue.

\subsection{Experimental Analysis}
\subsubsection{Impact of $\mathcal{L}_{\mathrm{Align}}$ on the Learned Transitions}
\begin{figure}[t]
     \centering
     \subfigure[Similarity between $\mathbf{h}_\mathrm{r2r}$ and $\mathbf{h}_\mathrm{s2r}$]{
        \label{fig:exp_his_cl_rec}
        \includegraphics[width=0.475\columnwidth]{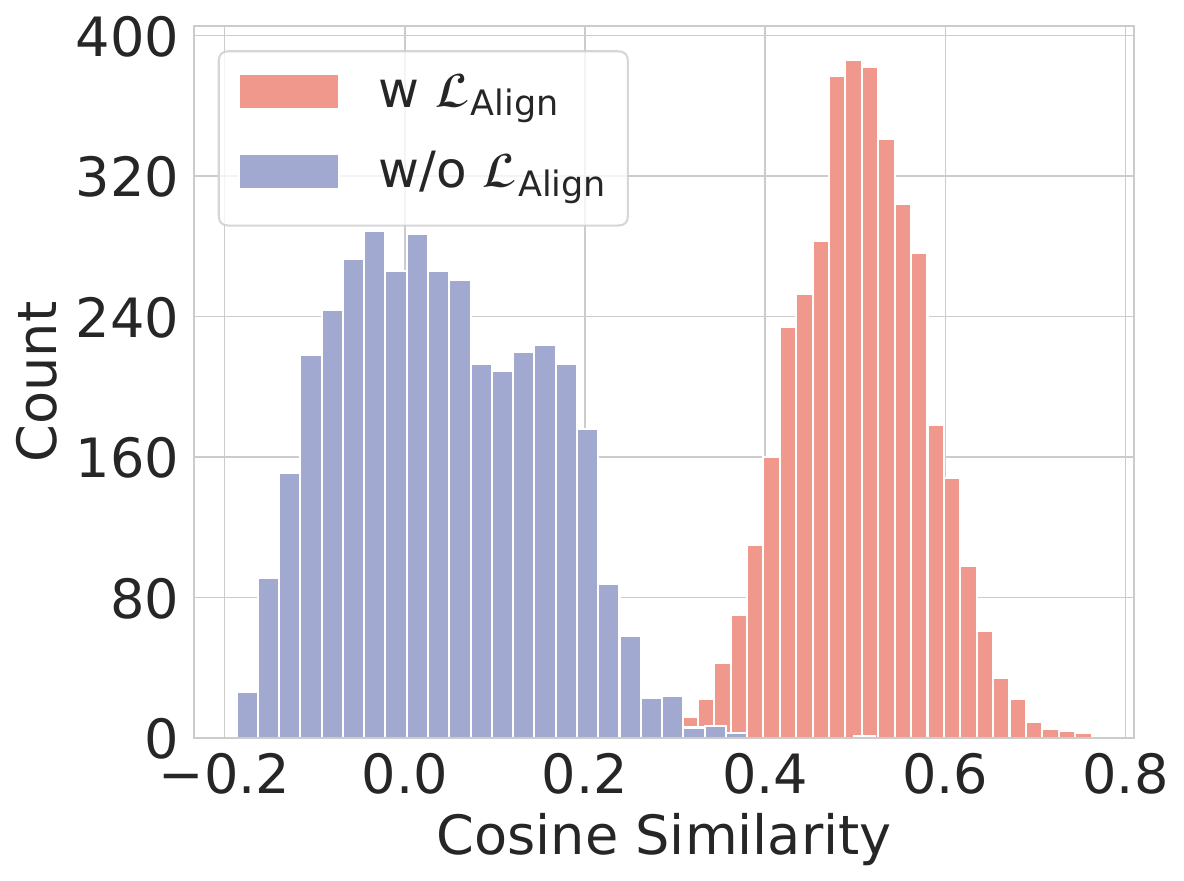}
     }
    \subfigure[Similarity between $\mathbf{h}_\mathrm{s2s}$ and $\mathbf{h}_\mathrm{r2s}$]{
        \label{fig:exp_his_cl_src}
        \includegraphics[width=0.475\columnwidth]{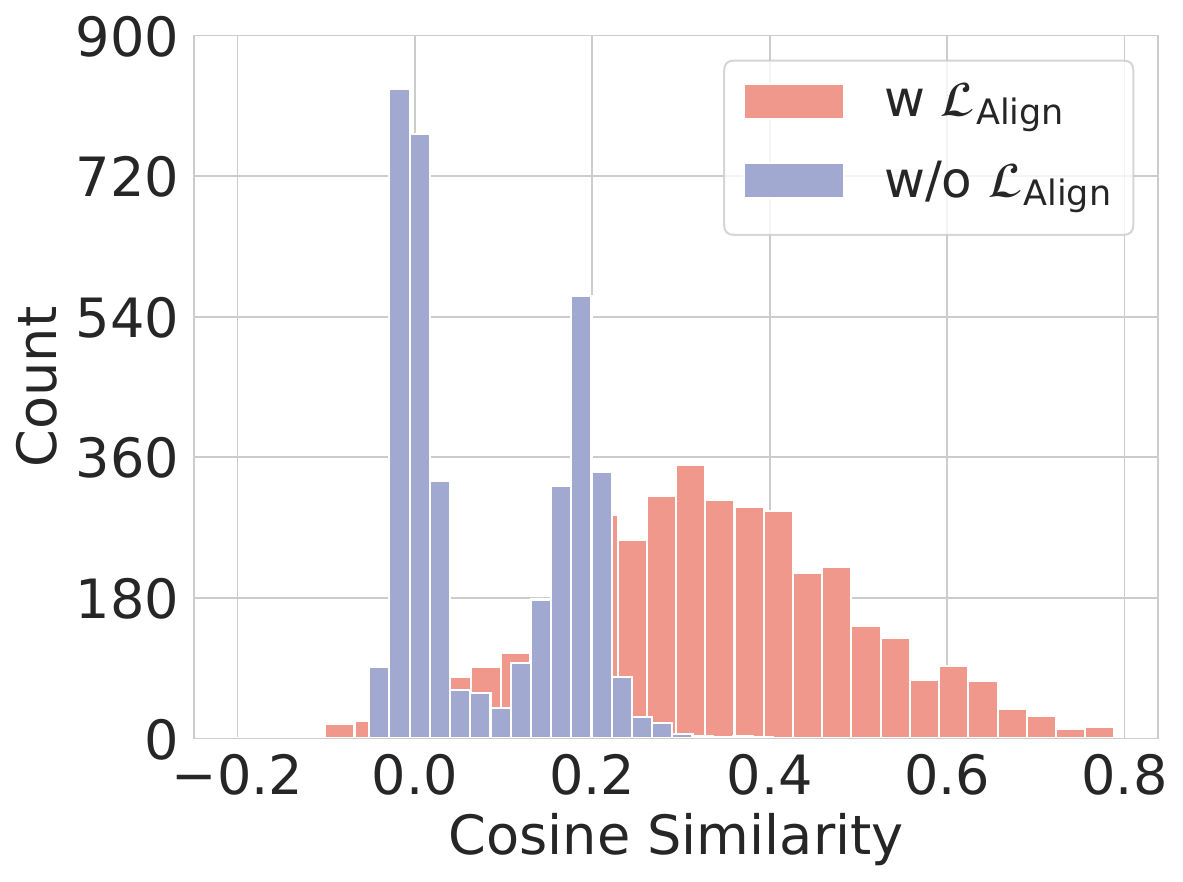}
     }
     \vspace{-5px}
     \caption{
     The histogram of the cosine similarities between the hidden representations, with and without introducing~$\mathcal{L}_{\mathrm{Align}}$.}
     \label{fig:exp_his_cl_sim}
     \vspace{-0.5cm}
\end{figure}

To further explore the impact of the contrastive learning in Section~\ref{sec:cl_before_fusion} on the learned representations of different transitions,
we computed the cosine similarity between $\mathbf{h}_{\mathrm{r2s}}$ and $\mathbf{h}_{\mathrm{s2s}}$, as well as between $\mathbf{h}_{\mathrm{s2r}}$ and $\mathbf{h}_{\mathrm{r2r}}$, with and without introducing the contrastive loss.
\autoref{fig:exp_his_cl_sim} depicts the distributions of the cosine similarities.
Without the loss $\mathcal{L}_{\mathrm{Align}}$, the similarity was primarily distributed around zero, indicating the lack of correlation between different transitions.
However, the similarity increased after integrating the contrastive loss $\mathcal{L}_{\mathrm{Align}}$.
This phenomenon indicates that $\mathcal{L}_{\mathrm{Align}}$ enables the model to capture similarities between different transitions.

\begin{figure}[t]
     \centering
     \subfigure[without $\mathcal{L}_{\mathrm{Align}}$]{
        \label{fig:exp_wo_his_cl}
        \includegraphics[width=0.475\columnwidth]{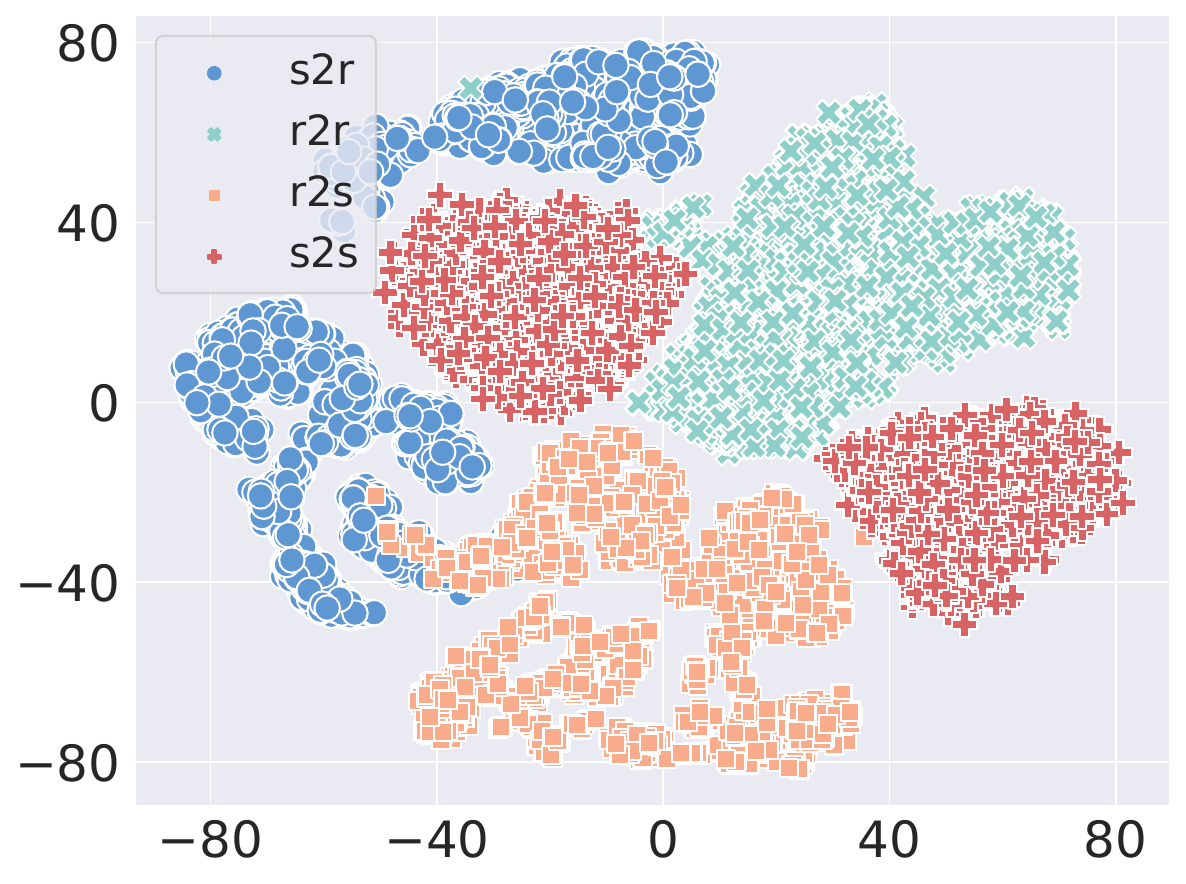}
     }
    \subfigure[with $\mathcal{L}_{\mathrm{Align}}$]{
        \label{fig:exp_w_his_cl}
        \includegraphics[width=0.475\columnwidth]{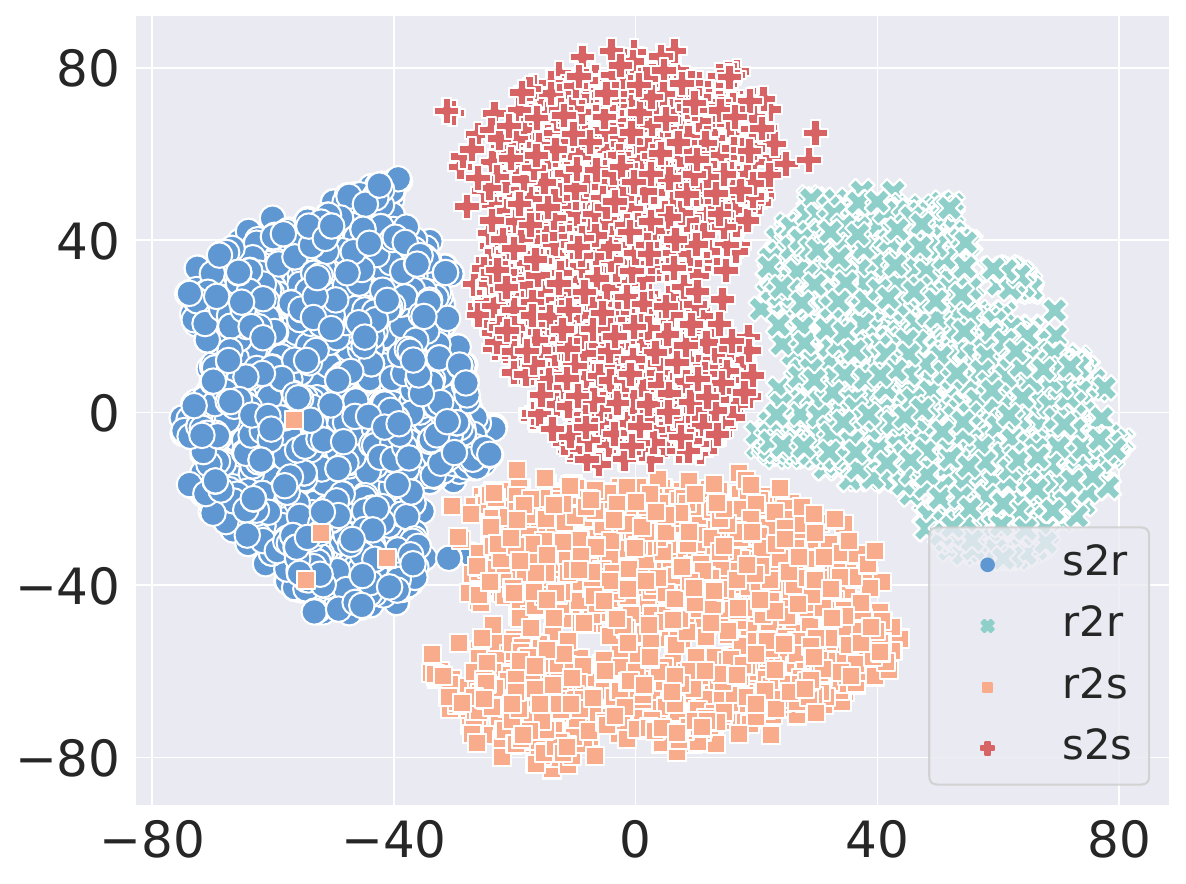}
     }
     \vspace{-5px}
     \caption{
     The t-SNE visualization of different transitions' hidden representations.}
     \label{fig:exp_his_cl_tsne}
     \vspace{-0.5cm}
\end{figure}

We additionally employed t-SNE~\cite{van2008visualizing} to visualize the representations of different transitions, as shown in Figure~\ref{fig:exp_his_cl_tsne}.
Without introducing the contrastive loss, we can observe that different transitions were relatively scattered, with transitions like \str and \sts unable to cluster effectively.
However, after incorporating the contrastive loss, various transitions were clustered more densely and compactly.
Specifically, representations of \str and \rtr became closer. Since \str and \rts originate from the same transformer, their representations tend to be closer. 
Overall, the inclusion of $\mathcal{L}_{\mathrm{Align}}$ facilitates a better understanding and learning of different transitions.

\subsubsection{Impact of Hyper-parameters}

\begin{figure}[t]
     \centering
     \subfigure[Performance of different $\alpha$]{
        \label{fig:exp_q_i_cl_weight}
        \includegraphics[width=0.475\columnwidth]{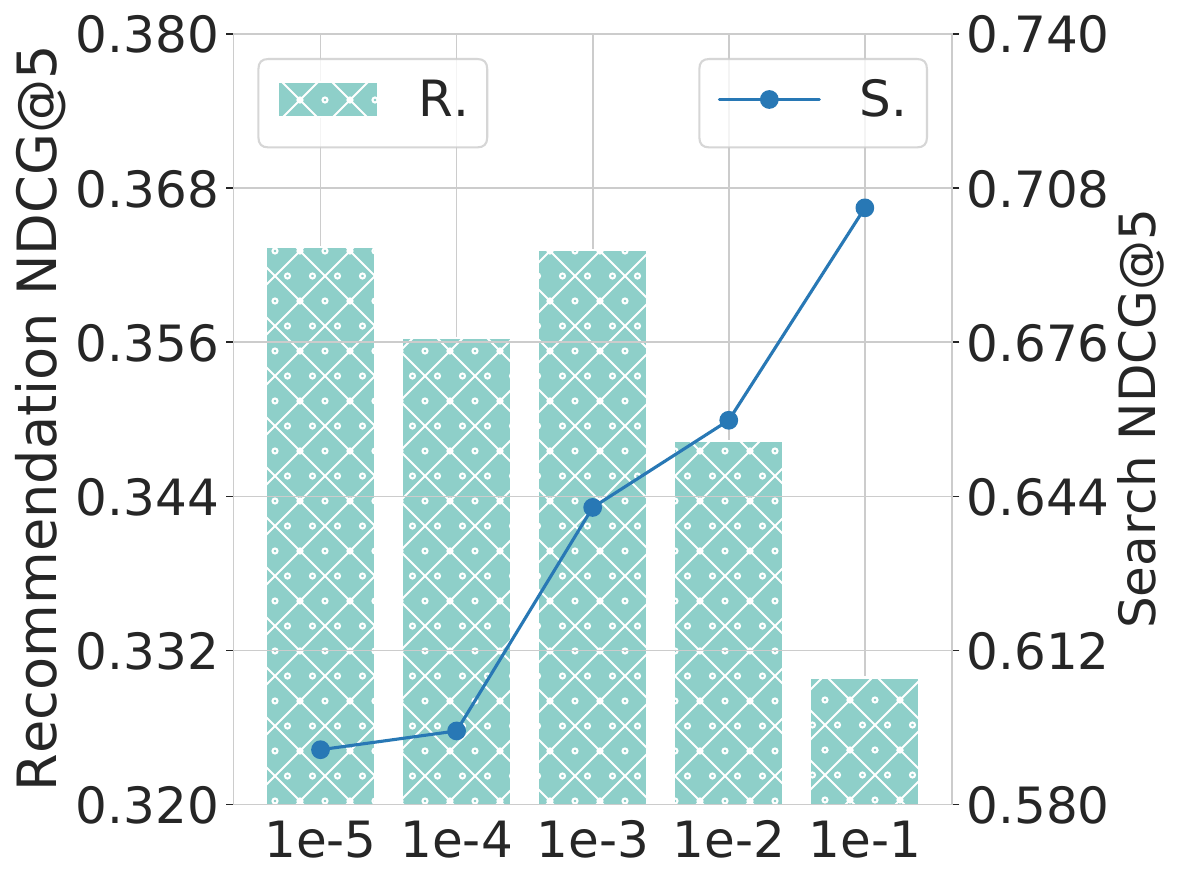}
     }
    \subfigure[Performance of different $\beta$]{
        \label{fig:exp_his_cl_weight}
        \includegraphics[width=0.475\columnwidth]{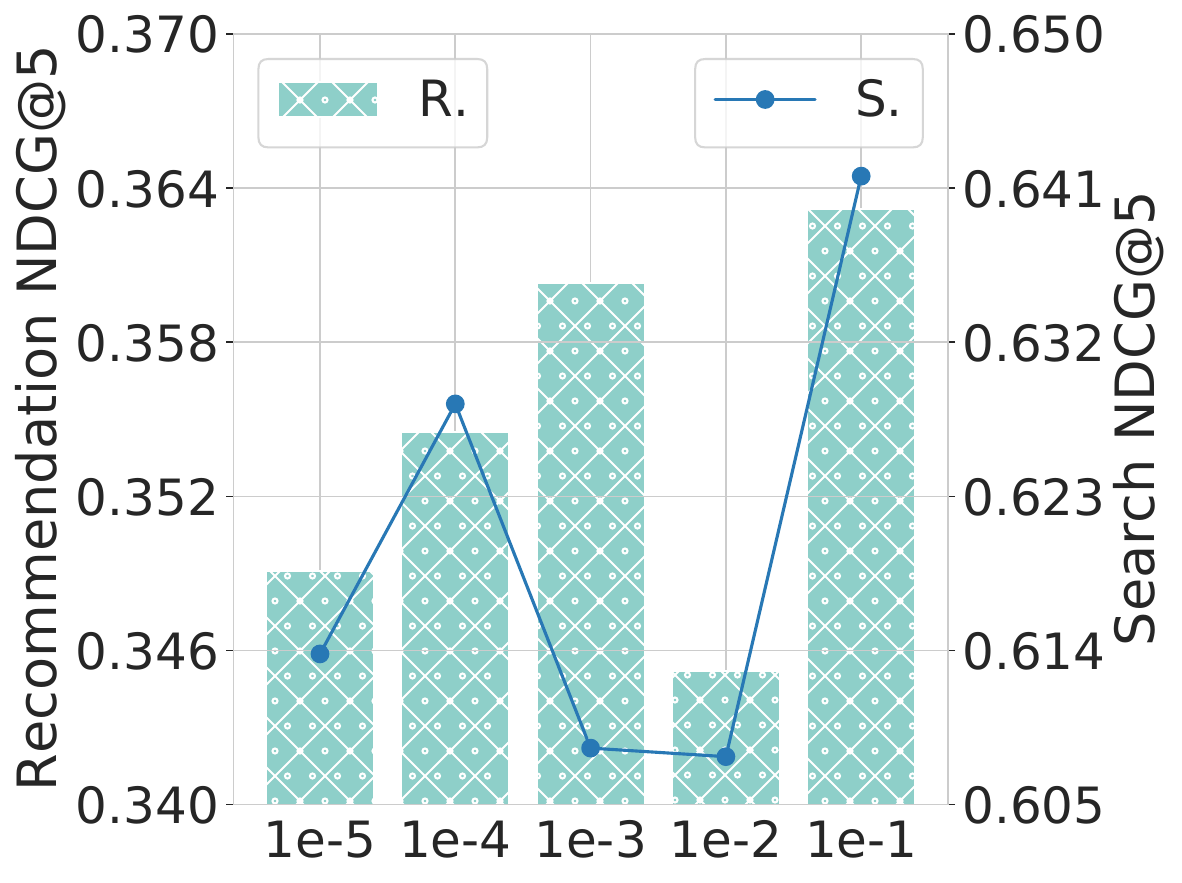}
     }
     \vspace{-5px}
     \caption{
     The recommendation and search performance with different values of $\alpha$ and $\beta$ in Eq.~\eqref{eq:loss_rec}}
     \label{fig:exp_cl_weight}
     \vspace{-0.5cm}
\end{figure}

We analyzed the effects of varying $\alpha$ and $\beta$ on the S\&R performance, as depicted in Figure~\ref{fig:exp_cl_weight}. 
While adjusting one parameter, the other remained constant at 1e-3 for $\alpha$ and 1e-1 for $\beta$.
Figure~\ref{fig:exp_q_i_cl_weight} illustrates that an increase in the weight $\alpha$ of $\mathcal{L}_{\mathrm{Rel}}$ corresponds to improved search performance but decreased recommendation performance. 
This is because $\mathcal{L}_{\mathrm{Rel}}$ tends to bring the query closer to its corresponding item, enabling the model to better learn their relevance. Search performance heavily relies on the relevance between queries and items.
Regarding recommendation, $\mathcal{L}_{\mathrm{Rel}}$ can augment performance to some extent. However, an excessively large $\alpha$ may hinder the optimization of click losses as defined in Eq.~\eqref{eq:loss_rec}, consequently leading to a decline in overall performance.
Concerning the weight $\beta$ of $\mathcal{L}_{\mathrm{Align}}$, optimal S\&R performance is achieved at a value of 1e-1.
When $\beta$ is too small, the model cannot effectively learn the similarity of different transitions, failing to efficiently fuse them.

\begin{figure}[t]
     \centering
     \subfigure[Recommendation Performance]{
        \label{fig:exp_mmoe_rec}
        \includegraphics[width=0.475\columnwidth]{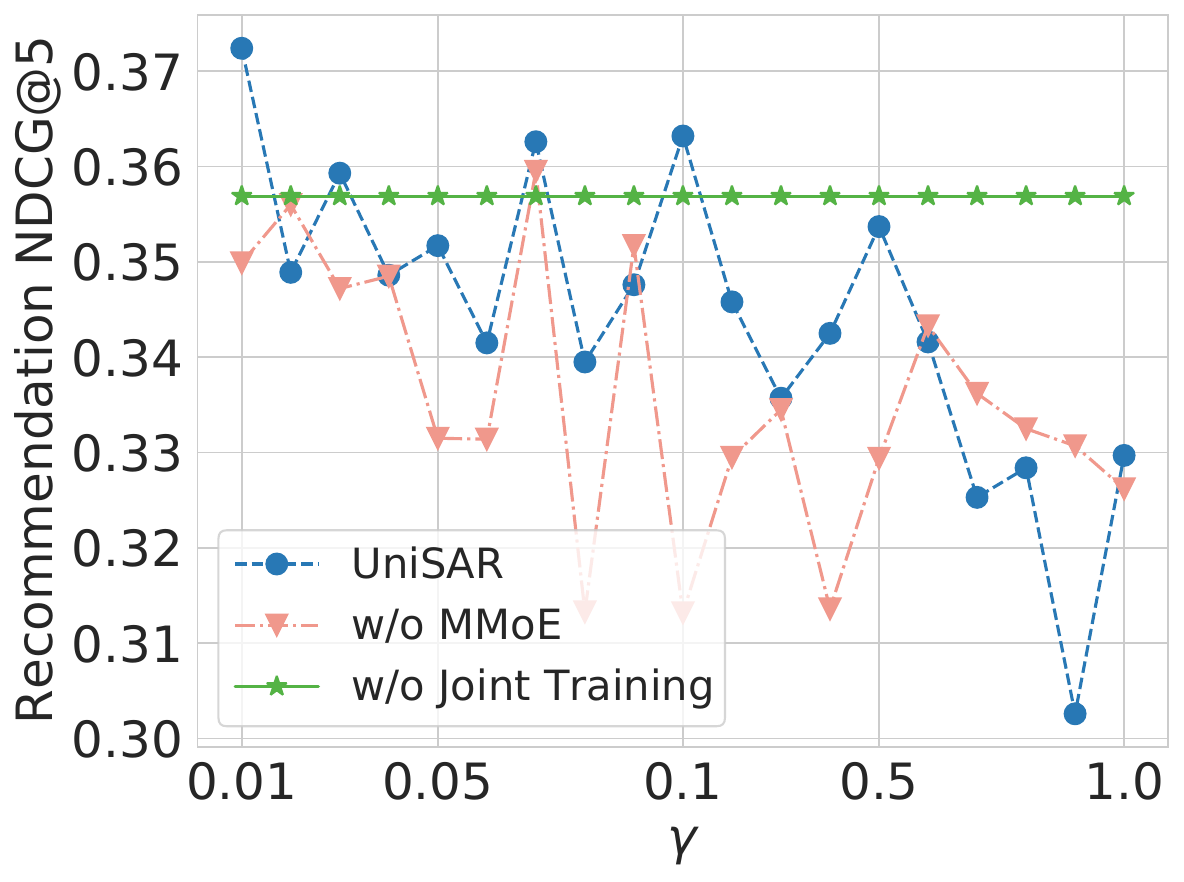}
     }
    \subfigure[Search Performance]{
        \label{fig:exp_mmoe_src}
        \includegraphics[width=0.475\columnwidth]{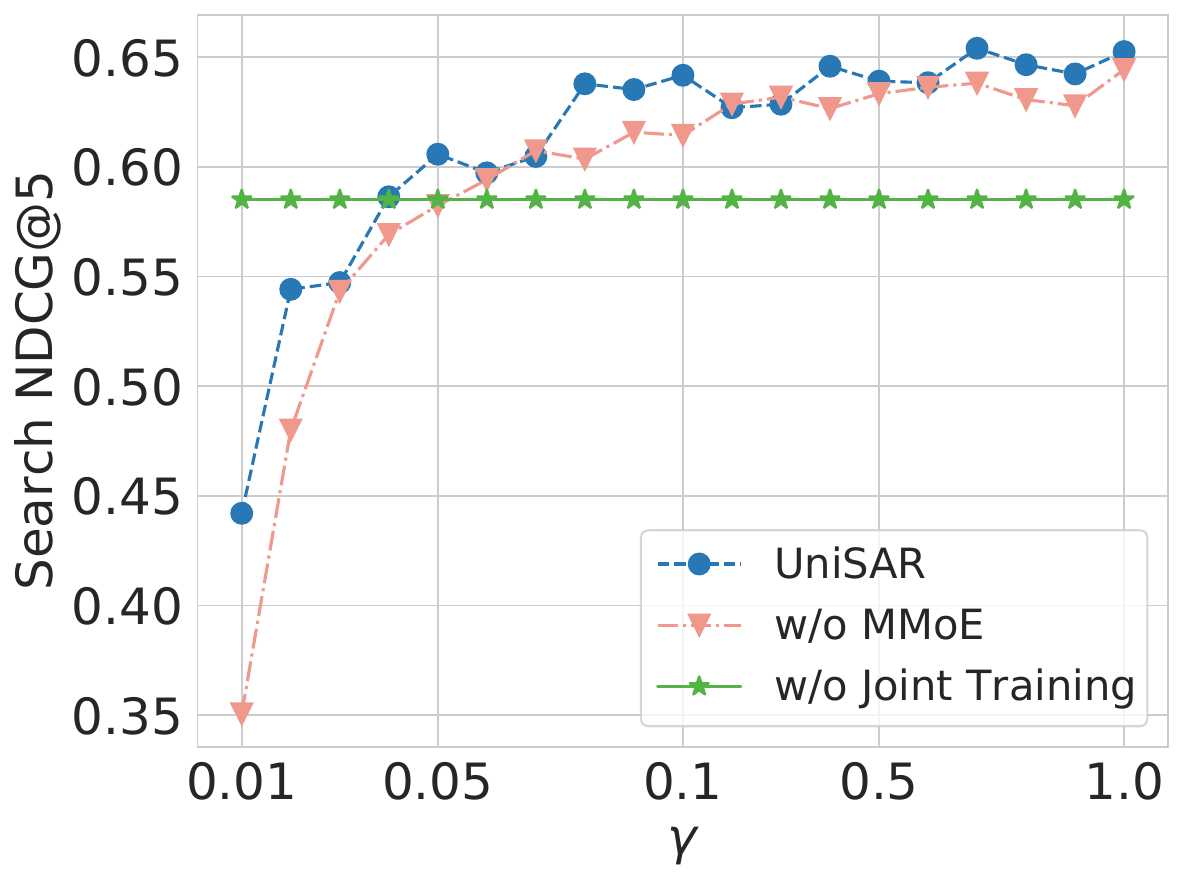}
     }
     \vspace{-5px}
     \caption{
     The recommendation and search performance with different values of $\gamma$ in Eq.~\eqref{eq:loss_total}}
     \label{fig:exp_mmoe_joint}
     \vspace{-0.5cm}
\end{figure}

We further investigated the impact of $\gamma$ in Eq.~\eqref{eq:loss_total} on the S\&R performance, as illustrated in Figure~\ref{fig:exp_mmoe_joint}.
It can be observed that without MMoE for joint training, an increase in $\gamma$ enhances search performance while diminishing recommendation effectiveness. Additionally, in this scenario, the recommendation performance falls below that of a model trained solely on recommendation data. This signifies the seesaw phenomenon between S\&R, highlighting the difficulty of effectively handling both tasks simultaneously through direct joint training without using MMoE.
With the introduction of MMoE, \ourname manages to achieve better results in most cases compared to models trained separately on S\&R data. This further demonstrates the significance of MMoE in learning shared and specific information between S\&R, thereby simultaneously enhancing the performance of both tasks.

\section{Conclusion}
In this paper, we propose \ourname which models fine-grained transitions between user S\&R behaviors. \ourname employs different transformers equipped with the mask mechanism to extract four transitions between S\&R and utilizes contrastive learning techniques to align them to learn the relationships between different transitions. 
Additionally, cross-attentions are employed to fuse different transitions to get the user representations.
\ourname is jointly trained on S\&R tasks using MMoE and can be applied to both scenarios, effectively utilizing knowledge from each to enhance the other.
Results on two public datasets validate the effectiveness of \ourname. 
The experimental analysis further illustrates the importance of each module within \ourname.

\begin{acks}
This work was funded by the National Key R\&D Program of China (2023YFA1008704), Kuaishou Technology, the National Natural Science Foundation of China (No. 62377044), Beijing Key Laboratory of Big Data Management and Analysis Methods, Major Innovation \& Planning Interdisciplinary Platform for the  ``Double-First Class” Initiative, Public Computing Cloud, funds for building world-class universities (disciplines) of Renmin University of China. 
\end{acks}

\clearpage
\bibliographystyle{ACM-Reference-Format}
\balance
\bibliography{ref}

\end{document}